\documentclass[11pt]{article}

\usepackage{amsmath,amssymb,amsfonts}
\usepackage{latexsym}
\usepackage{subcaption}
\usepackage{graphicx}
\usepackage{tabu}
\usepackage[backref=page]{hyperref}
\usepackage{xcolor}
\usepackage{wrapfig}
\usepackage{setspace}
\usepackage{xspace}

\usepackage[linesnumbered,lined,ruled,noend]{algorithm2e}
\usepackage[noend]{algpseudocode}
\usepackage{color}

\providecommand{\email}[1]{\href{mailto:#1}{\nolinkurl{#1}\xspace}}

\newcommand{\ff}{f}

\newenvironment{proof}{\noindent{\bf Proof : \ }}{\hfill$\Box$\par\medskip}

\newcommand{\coreset}{\textsc{Coreset}}
\newcommand{\basiccoreset}{\textsc{Basic\_Coreset}}

\renewcommand{\paragraph}[1]{\medskip\noindent\textbf{{#1}}}
\newcommand{\eps}{\ensuremath{\varepsilon}}                       
\newcommand{\br}[1]{\left\{#1\right\}}                            
\newcommand{\dist}{\mathrm{dist}}

\newcommand{\REAL}{\ensuremath{\mathbb{R}}}                       
\newcommand{\range}{\mathrm{range}}

\newcommand{\m}{m}

\newcommand{\ranges}{\mathrm{ranges}}

\newcommand{\pr}{\mathrm{pr}}

 \newtheorem{theorem}{Theorem}[section]
 
 \newtheorem{lemma}[theorem]{Lemma}
 
\newtheorem{definition}[theorem]{Definition}

 \newcommand{\CC}{Q}
\newcommand{\x}{x}
\newcommand{\E}{f}
\def\c{q}

\newcommand{\cost}{\ensuremath{\text{\footnotesize\textsf{COST}}}}

\newcommand{\PPPr}[2]{\ensuremath{\underset{#1}{\mathbf{Pr}}\left[#2\right]}}

\DeclareMathOperator*{\polylog}{polylog}
\newcommand{\sca}{0.31}

\usepackage{etoolbox}
\usepackage{fullpage}

\begin{document}
\title{New Frameworks for Offline and Streaming Coreset Constructions}
\date{\today}

\author{
Vladimir Braverman\thanks{Department of Computer Science, Johns Hopkins University. 
E-mail: \email{vova@cs.jhu.edu}}
\and
Dan Feldman\thanks{Computer Science Department, University of Haifa. 
E-mail: \email{dannyf.post@gmail.com}}
\and
Harry Lang\thanks{CSAIL, MIT. 
E-mail: \email{hlang08@gmail.com}}
\and
Adiel Statman\thanks{Computer Science Department, University of Haifa. 
E-mail: \email{statman.adiel@gmail.com}}
\and
Samson Zhou\thanks{School of Computer Science, Carnegie Mellon University. 
E-mail: \email{samsonzhou@gmail.com}}
}

\maketitle
\begin{abstract}
A coreset for a set of points is a small subset of weighted points that approximately preserves important properties of the original set. Specifically, if $P$ is a set of points, $Q$ is a set of queries, and $f:P\times Q\to\mathbb{R}$ is a cost function, then a set $S\subseteq P$ with weights $w:P\to[0,\infty)$ is an $\epsilon$-coreset for some parameter $\epsilon>0$ if $\sum_{s\in S}w(s)f(s,q)$ is a $(1+\epsilon)$ multiplicative approximation to $\sum_{p\in P}f(p,q)$ for all $q\in Q$. Coresets are used to solve fundamental problems in machine learning under various big data models of computation. Many of the suggested coresets in the recent decade used, or could have used a general framework for constructing coresets whose size depends quadratically on what is known as total sensitivity $t$. 

In this paper we improve this bound from $O(t^2)$ to $O(t\log t)$. Thus our results imply more space efficient solutions to a number of problems, including projective clustering, $k$-line clustering, and subspace approximation. Moreover, we generalize the notion of sensitivity sampling for sup-sampling that supports non-multiplicative approximations, negative cost functions and more. The main technical result is a generic reduction to the sample complexity of learning a class of functions with bounded VC dimension. We show that obtaining an $(\nu,\alpha)$-sample for this class of functions with appropriate parameters $\nu$ and $\alpha$ suffices to achieve space efficient $\epsilon$-coresets. 

Our result implies more efficient coreset constructions for a number of interesting problems in machine learning; we show applications to $k$-median/$k$-means, $k$-line clustering, $j$-subspace approximation, and the integer $(j,k)$-projective clustering problem. 
\end{abstract}

\section{Introduction}
Coresets are an important technique in machine learning, data sciences, and statistics for representing a large dataset with a much smaller amount of memory.  
Coresets are often used as a pre-processing dimensionality technique to improve the downstream efficiency of algorithms, both space and time. 
Informally speaking, a coreset $S$ of an input set $P$ of underlying points $p_1,\ldots,p_n$ is a smaller number of weighted representatives of $P$ that can be used to approximate the cost of any query from a set of a given queries. 
For example, in the common $k$-means clustering problem, the coreset must approximate $\sum_{i=1}^n d(p_i,C)^2$ for every query $C$, where $C$ is a set of $k$ points and $d(p_i,C)$ is taken to be the smallest Euclidean distance from $p_i$ to any point in $C$. 
Thus to use a coreset $S$ to approximately solve the $k$-means clustering problem, it suffices to find the optimal clustering on $S$ rather than find the optimal clustering on $P$. 
Because the size of $S$ is much smaller than the size of $P$, i.e., $|S|\ll|P|$, then finding an optimal clustering on $S$ instead of $P$ will be much more efficient. 

More generally, coreset is a set of points $P'$ with corresponding weight function $w(\cdot)$ such that $\sum_{p'_i\in P'} w(p')d(p'_i,C)^2$ is a $(1\pm\epsilon)$ approximation to $\sum_{i=1}^n d(p_i,C)^2$. 
Coresets have been extensively studied in $k$-means clustering~\cite{BadoiuHI02,Har-PeledM04,FrahlingS05,FrahlingS08,FeldmanL11,FeldmanS12,FeldmanSS13,BravermanLUZ19,HuangV20,FeldmanSS20}, subspace approximation~\cite{DeshpandeRVW06,DeshpandeV07,FeldmanL11,FeldmanMSW10,FeldmanSS13, ClarksonW15,SohlerW18}, and a number of other geometric problems and applications~\cite{AgarwalHY06,FeldmanFS06,Clarkson08,DasguptaDHKM08,AckermannB09,PhillipsT18,HuangJLW18,AssadiBBMS19,MunteanuSSW18,BravermanDMMUWZ18,MussayOBZF20}, due to the increasing availability of big data and the necessity for scalable methods to process this information. 

The most common algorithmic procedure to designing a coreset is the following simple template. 
An algorithm first approximately evaluates the \emph{sensitivity} of each point in the dataset. 
Informally, the sensitivity of a point quantities how important or distinct that point is, with respect to the given objective function on which we would like to optimize. 
Approximating the sensitivity of each point can often be done efficiently, so that the time to construct a coreset is often a lower order term compared to the runtime of the post-processing algorithm. 
The template then samples a fixed number of points, so that each point in the dataset with probability proportional to the sensitivity of the point. 
This approach is called sensitivity sampling and the fixed number of points is often a monotonically increasing function of the total sensitivity, defined to be the sum of the sensitivities of each point. 
Hence, if the numbered of sampled points is much smaller than the number of input points, this approach allows for compact dimensionality reduction, leading to improved performance of post-processing algorithms. 
 
Since the total sensitivity is a central quantity to coreset techniques, the total sensitivity for various objective functions has been well-studied and completely characterized in some cases. 
However, it is not quite known what the optimal dependency between the total sensitivity and the size of the coreset should be; that is, what is optimal monotonically increasing function of the total sensitivity that governs the number of sampled points? 
Clearly smaller functions lead to smaller coresets, which lead to more efficient post-processing functions. 
Many recent coreset constructions in the past decade require constructing coresets whose size depends quadratically on the total sensitivity. 
In this paper, we show this dependency is not optimal; we introduce in Theorem~\ref{thm:main} a generic construction whose dependency on the total sensitivity $t$ is only $O(t\log t)$ rather than $O(t^2)$~\cite{FeldmanL11}. 
Because the dependency is already black-boxed into the design of many coreset constructions, our results automatically improve many existing coreset algorithms simply by lowering the number of required samples, without modifying any other property of the algorithm; we are only showing that the worst-case theoretical guarantee of these algorithms is significantly and universally better than previously thought. 

\subsection{Our Contributions}
We show that the common sensitivity sampling framework only needs to sample $O(t\log t)$ points, where $t$ is the total sensitivity. 
\begin{theorem}
\label{thm:main}
Let $d$ be the dimension of a query space $(P,w,Q,f)$.
For each point $p$, let $m(p)$ be an upper bound on the sensitivity of point $p$. 
Let $t(p)=\sum_{p\in P}m(p)$, and $\eps,\delta\in(0,1)$. 
Then by sampling $O\left(\frac{t}{\eps^2}\left(d\log t+\log\left(\frac{1}{\delta}\right)\right)\right)$ i.i.d. points from $P$ and rescaling each sampled point by $\frac{1}{m(p)}$, the resulting sample is an $\epsilon$-coreset for $P$. 
\end{theorem}
In contrast, previous analysis showed that the sensitivity sampling framework required $O(t^2)$ points~\cite{FeldmanL11}. 
We emphasize that our results are purely theoretical; we show that any worst-case guarantee that could previously be achieved with $O(t^2)$ samples can actually be achieved with only $O(t\log t)$ samples. 
Hence our results can be universally plugged into any existing coreset construction algorithm simply by requiring a lower number of samples. 
Moreover, our results are optimal, since it can be shown by standard coupon-collector arguments that $\Omega(t\log t)$ samples are necessary in some cases. 

Our analysis is simple and uses results from the sample complexity of learning functions from unknown distributions~\cite{LiLS01}. 
Namely, let $X$ be a domain with an unknown underlying probability distribution $P$ and let $\mathcal{F}$ be a possibly infinite set of real-valued functionswith domain $X$. 
Suppose an algorithm is given independent samples $\x=(x_1,\ldots,x_m)$ from $P$ and oracle access to any $f\in\mathcal{F}$. 
Then how many samples must an algorithm observe before the sample average $\hat{\mathbb{E}}_{\x}(f):=\frac{1}{m}\sum_{i=1}^m f(x_i)$ is a rough approximation to $\mathbb{E}_{P}(f)=\underset{\x\sim P}{\mathbb{E}}f(\x)$? 
Note that since $\hat{\mathbb{E}}_{\x}(f)$ is an unbiased estimator for $\mathbb{E}_{P}(f)$, then standard concentration inequalities or variance bounding techniques demonstrate that the empirical average is roughly equal to the actual expectation. 
Surprisingly, \cite{LiLS01} proved that if $\mathcal{F}$ has pseudo-dimension $d$, then $O\left(\frac{1}{\alpha^2\nu}\log\frac{1}{\nu}\right)$ samples can \emph{simultaneously} obtain a good estimate to the expectation of all functions in $\mathcal{F}$ with constant probability. 

We show that the results of \cite{LiLS01} also succeed if the VC-dimension of $\mathcal{F}$ is $d$, rather than the pseudo-dimension. 
This implies that the algorithm outputs a $(p,\epsilon)$-approximation of all functions in $\mathcal{F}$, which means if the function is too small, then the resulting data structure can only provide an additive error guarantee rather than a multiplicative relative error guarantee, but if the function is adequately large, then the resulting data structure provides a multiplicative error guarantee. 

Fortunately, we show this guarantee suffices to obtain an $\eps$-coreset. 
We break the query space into partitions, based on how much a point contributes to a query, compared to the total contribution to a query across all the points. 
If the contribution of a partition is large, then our $(p,\epsilon)$-approximation guarantees a good approximation to this partition. 
Now if the contribution of a partition is small, then two things can happen. 
Either it is possible that the sum of the contributions of all of the ``small'' partitions is large, in which case our $(p,\epsilon)$-approximation again guarantees a good approximation, or the sum of the contributions remains insignificant. 
In this case, we only have an additive approximation of the contributions for these points, but because the sum of the contributions is insignificant, an additive error on these points translates to a small relative error on the entire objective function. 

\paragraph{Applications, Generalizations, and Empirical Evaluations.}
Theorem~\ref{thm:main} has applications to many problems in machine learning.  
In Section~\ref{sec:applications}, we describe applications to model fitting problems. 
Specifically, we consider the $(j,k)$-projective clustering problems such as $k$-median/$k$-means, $k$-line clustering, $j$-subspace approximation, and the integer $(j,k)$-projective clustering problem. 

Informally, the goal is to find a model $F$ in a restricted family $\mathcal{F}$ of set of $k$-tuples of affine $j$-subspaces that minimizes $\sum_{p\in P}d(p,F)$, where $P$ is a set of input points. 
Here, $F$ is the union of $k$ $j$-flats so that if $j=0$, then each $j$-flat reduces to a point and the $(j,k)$-projective clustering problem becomes the $k$-median problem with the appropriate metric. 
On the other hand, with an alternate distance function (the squared Euclidean distance), $d(\cdot,\cdot)$, the $(j,k)$-projective clustering objective becomes the $k$-means problem. 
When $j=1$ and $k$ is fixed, the objective becomes the $k$-line clustering problem but if $j$ is fixed and $k=1$, then the objective instead becomes the subspace approximation problem. 
Finally, in the integer $(j,k)$-projective clustering problem, all points in $P$ are assumed to have integer coordinates from some predetermined range. 
Our results subsume earlier versions online that have not received independent verification and are summarized in Figure~\ref{fig:applications}.

\begin{figure*}[!htb]
\begin{center}
\resizebox{\textwidth}{!}{
{\tabulinesep=1.2mm
\begin{tabular}{|c|c|c|c|}\hline
Problem & Coreset Size (Theorem~\ref{insofcor}) & Previous Coreset Size & Reference \\\hline
Integer $(j,k)$-projective clustering & $\tilde{O}\left(\frac{d}{\eps^2}g(d,j,k)(\log n)^{g(d,j,k)}\right)$ & $\tilde{O}\left(\frac{d}{\eps^2}g(d,j,k)(\log n)^{2\cdot g(d,j,k)}\right)$~\cite{FeldmanL11,VaradarajanX12a} & Thm.~\ref{thm:int:projective} \\\hline
$k$-line center & $\tilde{O}\left(\frac{d}{\eps^2}f(k)k^{f(k)}\log n\right)$ & $\tilde{O}\left(\frac{d}{\eps^2}k^{2\cdot f(k)}\log^2 n\right)$~\cite{FeldmanL11,VaradarajanX12b} & Thm.~\ref{thm:tight:kline} \\\hline
$k$-median & $O\left(\frac{d}{\eps^2}k\log k\right)$ & $O\left(\frac{d}{\eps^2}k^2\right)$~\cite{FeldmanL11,VaradarajanX12b} &  Thm.~\ref{thm:cluster} \\\hline
$k$-means & $O\left(\frac{d}{\eps^2}k\log k\right)$ & $O\left(\frac{d}{\eps^2}k^2\right)$~\cite{FeldmanL11,VaradarajanX12b} &  Thm.~\ref{thm:cluster} \\\hline
$j$-subspace fitting & $\tilde{O}\left(\frac{ds(j,d)}{\eps^2}\right)$ & $\tilde{O}\left(\frac{ds^2(j,d)}{\eps^2}\right)$~\cite{FeldmanL11,VaradarajanX12b} & Thm.~\ref{thm:fitting}\\\hline
\end{tabular}
}
}
\end{center}
\caption{Coreset sizes achieved by Theorem~\ref{insofcor} for various unsupervised learning problems. $f,g,s$ are functions independent of the input size $n$, specific to the problem setting. $\tilde{O}(\cdot)$ omits lower order terms.}
\label{fig:applications}
\end{figure*}


Although our primary contribution is theoretical, we complement our worst-case guarantees with empirical evaluations on both small and large-scale datasets, which we describe in Section~\ref{emp}. 

\subsection{Preliminaries}
\renewcommand{\pr}{\mathrm{Prob}}
For an integer $d\geq 1$, we denote by $\REAL^d=\REAL^{d\times 1}$ the set of column vectors in $\REAL^d$, and $[d]=\br{1,\cdots,d}$. 
For $\eps>0$, we denote by $1\pm\eps$ the interval $[1-\eps,1+\eps]$. 
A multiplication of a real number $c\in\REAL$ by a set $X\subseteq\REAL^d$ is denoted by $cX=\br{cx\mid x\in \REAL^d}$. 
A sum or minimum over an empty set is defined to be $0$ in this paper.
We say a set of random variables are i.i.d. if the random variables are independent and identically distributed. 
We use $\tilde{O}(\cdot)$ to suppress lower order terms, e.g. $\tilde{O}\left(\frac{1}{\eps^2}\log n\right)$ omits $\polylog\left(\frac{1}{\eps}\right)$ and $\polylog\log n$ terms. 

\begin{definition}[Weighted Set]
Let $X$ be called a \emph{ground set}. 
Let $P\subseteq X$ be a (possibly ordered) multi-set and $w:P\to[0,\infty)$ be a function that maps every $p\in P$ to a weight $w(p)\geq 0$. 
The pair $(P,w)$ is called a \emph{weighted set} in $X$. 
If $w(p)=1$ for every $p\in P$ then the (un)weighted set $(P,w)$ may be denoted by $P$ for short.
\end{definition}
The order of the points in $P$ can be arbitrary in this paper. 
However, even if $P$ contains only a single copy of each point, the corresponding coreset may contain multiple instances of some points. 
Hence, we consider coresets as multi-sets, although duplicated points can usually be replaced by a single weighted point without changing the claimed results.
The union and intersection are also implied to be over multi-sets in this paper. 

\begin{definition}[Query and Range Space]\label{def::query space}
Let $X$ be a ground set and $(P,w)$ be a weighted set in $X$ called the \emph{input set}. 
Let $\CC$ be a \emph{query function} that maps every set $S\subseteq X$ to a corresponding $\CC(S)$, such that $\CC(T)\subseteq \CC(S)$ for every $T\subseteq S$. 
Let $\ff:X\times \CC(X)\to \REAL$ be called a \emph{loss function}.
The tuple $(P,w,\CC,\ff)$ is called a \emph{query space}. 
For a collection $\mathcal{R}$ of subsets of $X$, we call $(X,\mathcal{R})$ a \emph{range space}. 
For every weighted set $C'=(C,u)$ in $X$, and every $q\in \CC$ we define the overall fitting error of $C'$ to $q$ by
$f(C',q)=\sum_{p\in C}u(p)f(p,q)$.
\end{definition}
For example, in the $k$-means clustering problem on $\mathbb{R}^n$, the ground set $X$ is the domain $\mathbb{R}^n$ that contains is the set $P$ of input points. 
The query function for $k$-means clustering restricts queries of $\mathbb{R}^n$ to $k$ points and the loss function is the squared Euclidean distances, so that when the arguments of the loss function are a point from $P$ and a point from the query function, the loss function is the squared distance from the input point to the closest center. 

Usually the loss function has specific properties such as being a pseudo distance function $D$, as will be defined later. 
However, it may also be more complicated such as a subtraction between pseudo distance functions, which will also be used in this paper. 
This is also why it may return a negative number. 
In general, we will be interested in approximating $\sum_{p\in P}w(p)f(p,q)$ for every query $q$ in the query space up to an additive error of $\eps$. 

For a set $X$, a query function $\CC$, and a cost function $\ff$, we define the VC-dimension of the range space that it induced, as defined below. The classic VC-dimension was defined for sets and subset and here we generalize it to query spaces, following~\cite{FeldmanL11}.
\begin{definition}[VC-dimension]
\cite{VapnikC71,BravermanFL16,FeldmanL11}
\label{vdim}
For a ground set $X$ and a set $\ranges$ of subsets of $X$, the VC-dimension of $(X,\ranges)$ is the size $|C|$ of the largest subset $C\subseteq X$ such that
\[|\br{C\cap \range \mid \range\in\ranges}|= 2^{|C|}.\]
Let $Q$ be a query function and $f:X\times Q(X)\to\REAL$.
For every $q\in \CC(X)$, and $r\in\REAL$ we define the sets
\begin{align*}
\range_{P,\ff}(\c,r)&:=\br{p\in P\mid \ff(p,\c)\leq r},\\
\ranges(P,\CC,f)&:=\{C\cap \range_{P,\ff}(\c,r)\mid C\subseteq P,\\
&\qquad \c\in \CC(C), r\in\REAL\}.
\end{align*}
The \emph{dimension} of $(P,\CC,\ff)$ is the VC-dimension of $(P,\ranges(P,\CC,f))$.
\end{definition}

The following definition of sensitivity is central to our paper, as we shall show that the coreset size of our algorithm is proportional to the total sensitivity of the input set. 
\begin{definition}[Sensitivity]
Let $(P,w,\CC,\E)$ be a query space over a ground set $X$, where $w:P\to [0,\infty)$ and $f:P\times Q(P)\to[0,\infty)$. 
Then we define the \emph{sensitivity} of a point $p\in P$ by $s(p)=\sup_{C\in\CC}w(p)|f(p,C)|$. 
\end{definition}
Then the total sensitivity of an input set is the natural definition:
\begin{definition}[Total Sensitivity]
Let $(P,w,\CC,\E)$ be a query space over a ground set $X$, where $w:P\to [0,\infty)$ and $f:P\times Q(P)\to[0,\infty)$. 
We define the \emph{total sensitivity} of $P$ by $\sum_{p\in P}s(p)$, where $s(p)$ is the sensitivity of $p$.  
\end{definition}
We next define two related concepts, the $(\nu,\alpha)$-samples and relative $(p,\eps)$-approximations. 
\begin{definition}[$(\nu,\alpha)$-Sample]
\cite{LiLS01}
Let $\alpha,\nu>0$. 
For every $a,b\geq 0$, we define the distance function $d_{\nu}(a,b)=\frac{|a-b|}{a+b+\nu}$.
Let $(P,w,\CC,\E)$ be a query space over a ground set $X$, where $w:P\to [0,\infty)$ and $f:P\times Q(P)\to[0,\infty)$. 
Then the weighted set $(S,u)$ is called a \emph{$(\nu,\alpha)$-sample} for $(P,w,\CC,\E)$ if $(S,u,\CC,\E)$ is a query space, and for every $q\subseteq\CC(S)$, $d_{\nu}(\overline{\E}(P,w,;\c),\overline{\E}(S,u,\c))\leq\alpha$, where $\overline{\E}(P,w,\c)=\frac{\sum_{p\in P}|w(p)\cdot\E(p,\c)|}{\sum_{p\in P}w(p)}$. 
\end{definition}

\begin{definition}[$(p,\eps)$-Approximation]
\cite{Har-PeledS11}
Let $0<p,\eps<1$. 
Let $(P,w,\CC,\E)$ be a query space over a ground set $X$, where $w:P\to [0,\infty)$ and $f:P\times Q(P)\to[0,\infty)$. 
Then the weighted set $(S,u)$ is called a \emph{$(\nu,\alpha)$-sample} for $(P,w,\CC,\E)$ if $(S,u,\CC,\E)$ is a query space, and for every $q\subseteq\CC(S)$,
\begin{enumerate}
\item
$(1-\eps)\overline{\E}(P,w,\c)\le\overline{\E}(S,u,\c)\le(1+\eps)\overline{\E}(P,w,\c)$, for $\overline{\E}(P,w,\c)\ge p$
\item
$\overline{\E}(P,w,\c)-\eps p\le\overline{\E}(S,u,\c)\le\overline{\E}(P,w,\c)+\eps p$, for $\overline{\E}(P,w,\c)\le p$,
\end{enumerate}
where $\overline{\E}(P,w,\c):=\frac{\sum_{p\in P}|w(p)\cdot\E(p,\c)|}{\sum_{p\in P}w(p)}$. 
\end{definition}
We recall the equivalence between $(\nu,\alpha)$-samples and relative $(p,\eps)$-approximations:
\begin{theorem}
\cite{Har-PeledS11}
\label{sample:approx:same}
Let $(X,\mathcal{R})$ be a range space. 
If $(Z,\mathcal{R})$ is a $(\nu,\alpha)$-sample for $(X,\mathcal{R})$ with $0<\alpha<\frac{1}{4}$ and $\nu>0$, then $Z$ is a relative $(\nu,4\alpha)$-approximation for $(X,\mathcal{R})$. 
\end{theorem}
\begin{definition}[$\eps$-coreset]
Let $P'=(P,w)$ be a weighted set in $X$, and $\eps> 0$ be an approximation error. The weighted set $C'=(C,u)$ in $X$ is an \emph{$\eps$-coreset} for a query space $(P',\CC,f)$ if for every $q\in \CC(C)$ we have $(1-\eps)\sum_{p\in C}u(p)\ff(p,q)\le\sum_{p\in P}w(p)\ff(p,q)\\
\le(1+\eps)\sum_{p\in C}u(p)\ff(p,q)$.
\end{definition}
\section{Sensitivity Sampling}
\label{sec:sample}
In this section, we show that provable worst-case guarantees for constant factor approximation can be achieved using the sensitivity sampling framework to construct coresets of size $O(t\log t)$, where $t$ is the total sensitivity\footnote{We also achieve optimal dependence on $\frac{1}{\epsilon}$ for a $(1+\epsilon)$-approximation, but we omit these factors for ease of discussion.} 
This improves on previous analysis that the sensitivity sampling framework to sample $O(t^2)$ points to construct coresets that guaranteed constant factor approximation~\cite{FeldmanL11}. 
We remark that our result is purely theoretical and does not require novel algorithmic implementation. 
Instead, our result shows that the parameters in existing coreset construction algorithms can be improved while still guaranteeing worst-case performance. 

Recall that the sensitivity of a point $p\in P$ is defined by $s(p)=\sup_{C\in Q(P)}w(p)|f(p,C)|$, where $w:P\to [0,\infty)$ is a weight function $p$ and $f:P\times Q(P)\to[0,\infty)$ is a loss function between an input point and a query set.  
However, determining the exact sensitivity of a point can be time-consuming, so we instead define $m(p)\ge s(p)$ to be an upper bound on the sensitivity. 
It turns out that upper bounds $m(p)$ that are within a constant factor approximation of the exact sensitivity of a point are often efficiently computable. 
Then $t:=t(P)=\sum_{p\in P} m(p)$ is an upper bound on the total sensitivity, which is the sum of the sensitivities of all points in $P$. 

We now formalize the sensitivity sampling framework broadly used in algorithmic design. 
We form a sample $S$ by picking the first point of $S$ to be $p\in P$ with probability $\frac{m(p)}{t(P)}$ and reweighting the sampled point with the inverse of the sampling probability. 
We show that repeatedly sampling points from $P$ with replacement until $S$ has $\frac{ct}{\eps^2}\left(d\log t+\log\left(\frac{1}{\delta}\right)\right)$ points suffices to obtain an $\eps$-coreset for $P$ with probability $1-\delta$ if the underlying query space has VC dimension $d$. 
The sensitivity sampling framework appears in full in Algorithm~\ref{alg:coreset}.

\setcounter{algocf}{-1}
\begin{algorithm}
\caption{{\basiccoreset}($P,d,\eps,\delta$)}
\label{alg:coreset}
\textbf{Input:} Set of $n$ points $P$ in a query space with VC dimension $d$, approximation parameter $\eps>0$, failure probability $\delta>0$, oracle access to upper bound $m(\cdot)$ on sensitivities.\\
\textbf{Output:} $\eps$-coreset of $P$.
$t\gets 0$\\
\For{every $p\in P$}
{Let $m(p)$ be an upper bound on the sensitivity $s(p)$ of $p$.\\
$t\gets t+m(p)$
}
$S\gets\emptyset$\\
$N\gets\frac{ct}{\eps^2}\left(d\log t+\log\left(\frac{1}{\delta}\right)\right)$ for sufficiently large constant $c>0$ that can be determined from the proof.\\
\For{$i=1$ to $N$}
{With probability $\frac{m(p)}{t}$, set $x=p$ for $p\in P$ with weight $\frac{t}{m(p)}$.\\
$S\gets S\cup\{x\}$
}
\Return $S$
\end{algorithm}

\subsection{Sample Complexity of Learning}
We first consider the problem of relating the sample complexity of learning a class of functions to its VC dimension. 
We then show this implies an $\eps$-coreset construction under certain parameters. 
Let $X$ be a domain with probability distribution $\mu$ and let $\mathcal{F}$ be a possibly infinite set of real-valued functions defined on $X$. 
Given access to samples $\x=(x_1,\ldots,x_N)$ independently drawn from $\mu$ and oracle access to any $f\in\mathcal{F}$, the sample average $\hat{\mathbb{E}}_{\x}(f):=\frac{1}{N}\sum_{i=1}^N f(x_i)$ serves as an unbiased estimator to the expectation of $f$, denoted $\mathbb{E}_{\mu}(f)=\underset{\x\sim\mu}{\mathbb{E}}f(\x)$. 
For sufficiently large $N$, standard concentration inequalities can show that the sample average $\hat{\mathbb{E}}_{\x}(f)$ is an $\alpha$-approximation to the expectation $\mathbb{E}_{\mu}(f)$, where $0<\alpha\le 1$. 
However, it is not obvious whether there exists a value of $N$ that allows simultaneous estimation of the expectation of \emph{all} functions in $\mathcal{F}$. 

We first recall the following definition of pseudo-dimension:
\begin{definition}[Pseudo-dimension]
The \emph{pseudo-dimension} of a class $\mathcal{F}$ of functions from a domain $X$ to $[0,1]$ is defined to be the largest $d$ such that there exists a sequence $x_1,\ldots,x_d\in X$ and a sequence $r_1,\ldots,r_d\in\mathbb{R}$ of thresholds such that for all $2^d$ combinations of $b_1,\ldots,b_d\in\{0,1\}$, there exists an $f\in\mathcal{F}$ such that for all $i\in[d]$, $f(x_i)\ge r_i$ if and only if $b_i=1$. 
\end{definition}

\cite{LiLS01} show that if $\mathcal{F}$ has pseudo-dimension $d$, then $O\left(\frac{1}{\alpha^2\nu}\log\frac{1}{\nu}\right)$ samples suffices to simultaneously obtain an $(\nu,\alpha)$-sample to expectation of all functions in $\mathcal{F}$ with constant probability. 
Namely, \cite{LiLS01} show the following two lemmas:
\begin{lemma}[Lemma 6 in \cite{LiLS01}]
\label{lls:lem:six}
Let $\mathcal{F}$ be a set of functions from $X$ to $[0,1]$, $\mu$ be a probability distribution over $X$ and $\nu>0$, $0<\alpha<1$, and $N\ge\frac{2}{\alpha^2\nu}$. 
For any integer $N>0$, let $\Gamma_N$ denote the set of all permutations of $\{1,\ldots,2N\}$ so that for each $i\le N$, either $i$ and $N+i$ are fixed, or $i$ and $N+i$ are swapped. 
Let $U$ be the uniform distribution over $\Gamma_m$. 
Then
\begin{align*}
&\PPPr{N}{\x\,:\,\exists f\in\mathcal{F}, d_{\nu}(\hat{\mathbb{E}}_{\x}(f),\mathbb{E}_{\mu}(f))>\alpha}\le 2\underset{\x\in\mathbb{X}^{2N}}{\sup} U\Big\{\sigma\,:\,\\
&\exists f\in\mathcal{F}, d_{\nu}\Big(\frac{1}{N}\sum_{i=1}^Nf(x_{\sigma(i)}),\frac{1}{N}\sum_{i=1}^Nf(x_{\sigma(Nn+i)})\Big)>\frac{\alpha}{2}\Big\}.
\end{align*}
\end{lemma}

\begin{lemma}[Lemma 10 in \cite{LiLS01}]
\label{lls:lem:ten}
Let $d$ be the pseudo-dimension of $F\subseteq[0,1]^{2N}$, where $N\ge\frac{125(2d+1)}{\alpha^2\nu}$ for any $\alpha,\nu>0$. 
Let $U$ be the uniform distribution over $\Gamma_N$. 
Then
\begin{align*}
U\{\sigma\,:\,\exists f\in F,d_{\nu}(\mu_1(f,\sigma),\mu_2(f,\sigma))>\alpha\}\le6\left(\frac{2624}{\nu}\right)^d e^{-\frac{\alpha^2\nu N}{90}}.
\end{align*}
\end{lemma}
Observe that combining Lemma~\ref{lls:lem:six} and Lemma~\ref{lls:lem:ten} and solving for $N$ recovers the bound from \cite{LiLS01} of $O\left(\frac{1}{\alpha^2\nu}\left(d\log\frac{1}{\nu}+\log\frac{1}{\delta}\right)\right)$. 
We need an analog of their sampling result for VC-dimension rather than pseudo-dimension. 
As it turns out, the only place \cite{LiLS01} uses pseudo-dimension in Lemma~\ref{lls:lem:ten} is a black-box reduction from the following lemma to bound the size of $\mathcal{F}$:
\begin{lemma}
\cite{Haussler95,LiLS01}
\label{lls:lem:nine}
For $v,w\in\mathbb{R}^k$, let $\ell_1(v,w)=\frac{1}{k}|v_i-w_i|$. 
Let $k>0$ be an integer and $0<\alpha\le$. 
Suppose each pair of distinct elements $f,g\in F\subseteq[0,1]^k$ has $\ell_1(f,g)\le\alpha$. 
Then $|F|\le\left(\frac{41}{\alpha}\right)^d$, where $d$ is the pseudo-dimension of $F$. 
\end{lemma}
Moreover, \cite{Haussler95} proved the exact same statement when $d$ is the VC-dimension of $F$, rather than the pseudo-dimension. 
\begin{lemma}
\cite{Haussler95}
\label{lls:replacement}
For $v,w\in\mathbb{R}^k$, let $\ell_1(v,w)=\frac{1}{k}|v_i-w_i|$. 
Let $k>0$ be an integer and $0<\alpha\le$. 
Suppose each pair of distinct elements $f,g\in F\subseteq[0,1]^k$ has $\ell_1(f,g)\le\alpha$. 
Then $|F|\le\left(\frac{41}{\alpha}\right)^d$, where $d$ is the VC-dimension of $F$. 
\end{lemma}
Specifically, Lemma~\ref{lls:replacement} follows from Corollary 1 in~\cite{Haussler95} because $e(d+1)(2e/\eps)^d\le(2e^3/\eps)^d<(41/\eps)^d$.

Thus by using Lemma~\ref{lls:replacement} rather than Lemma~\ref{lls:lem:nine}, we can recover Lemma~\ref{lls:lem:ten} using VC-dimension rather than pseudo-dimension in the following formulation of Theorem~\ref{thm:lls:sample}. 
Hence, we can relate the sampling complexity of learning a class of functions to their VC-dimension:
\begin{theorem}
\label{thm:lls:sample}
Let $\delta\in(0,1)$ be a failure probability and $\mathcal{F}:X\to[0,1]$ be a family of functions with VC dimension $d$. 
Then with probability at least $1-\delta$, $N=O\left(\frac{1}{\alpha^2\nu}\left(d\log\frac{1}{\nu}+\log\frac{1}{\delta}\right)\right)$ samples suffice to simultaneously obtain an $(\nu,\alpha)$-sample to any $f\in\mathcal{F}$, i.e., $d_{\nu}(\hat{\mathbb{E}}_{\x}(f),\mathbb{E}_{\mu}(f))<\alpha$. 
\end{theorem}

\subsection{Reduction to $\eps$-Coresets}
We now show that an $(\nu,\alpha)$-sample to a class of functions $\mathcal{F}$ suffices to achieve an $\eps$-coreset under the appropriate parameters. 
The proof partitions the points in an input set $P$ by their contribution to $\sum_{p\in P}w(p)f(p,X)$ for some $X$ in the query space. 
A subset $S_i$ that contributes a large fraction towards $\sum_{p\in P}w(p)f(p,X)$ will be well-estimated by the $(\nu,\alpha)$-sample. 
On the other hand, if $S_i$ is not well-estimated by the $(\nu,\alpha)$-sample, then its contribution towards $\sum_{p\in P}w(p)f(p,X)$ must be small, so that intuitively, the additive error from the $(\nu,\alpha)$-sample is also small. 
Thus we can show that the sample is actually an $\eps$-coreset.

\begin{theorem}[Theorem~\ref{thm:main}, Restated]\label{insofcor}
Let $d$ be the dimension of a query space $(P,w,Q,f)$.
Suppose that $m:P\to[0,\infty)$ such that $m(p)\geq \sup_{C\in Q(P)}w(p)|f(p,C)|$.
Let $t\geq \sum_{p\in P}m(p)$, and $\eps,\delta\in(0,1)$. 
Let $c\geq1$ be a sufficiently large constant, and let $S$ be a sample of
\[
|S|\geq \frac{ct}{\eps^2}\left(d\log t+\log\left(\frac{1}{\delta}\right)\right)
\]
i.i.d. points from $P$, where for every $p\in P$ and $s\in S$ we have $\pr(p=s)\geq\frac{m(p)}{t}$.
Then, with probability at least $1-\delta$, then simultaneously for all $C\in Q(S)$, 
\[
\left|\sum_{p\in P} w(p)f(p,C)-\sum_{q\in S} \frac{w(q)}{|S|\pr(q)}\cdot f(q,C)\right| \leq \eps.
\]
\end{theorem}
\begin{proof}
Let $(Z,\mathcal{R})$ be a $\left(\frac{1}{t},\eps\right)$-sample for $(X,\mathcal{R})$. 
By Theorem~\ref{sample:approx:same}, $Z$ is a $\left(\frac{1}{t},\eps\right)$-approximation for $(X,\mathcal{R})$. 
Let $w(p)$ be the weight of a point $p\in P$ given by $X$ and $u(p)$ be the weight of $p$ given by $Z$. 
We partition of the range space into sets $S_i$, where $i\in\mathbb{Z}$, so that $S_i=\{p\in P\,:\,2^i\le w(p)f(p,X)<2^{i+1}\}$. 

Let $\mu_i=\frac{\sum_{p\in S_i}w(p)f(p,X)}{\sum_{p\in P}w(p)f(p,X)}$ be the fraction of the total mass that $S_i$ contributes. 
Let $A_1=\left\{i\in\mathbb{Z}\,:\,\mu_i<\frac{1}{t}\right\}$ be the set of indices $i$ such that $S_i$ contributes less than $\frac{1}{t}$ fraction of the mass and let $P_1=\cup_{i\in A_1}S_i$ be the points in $A_1$. 
Similarly, let $A_2=\left\{i\in\mathbb{Z}\,:\,\mu_i\ge\frac{1}{t}\right\}$ be the set of indices $i$ such that $S_i$ contributes at least $\frac{1}{t}$ fraction of the mass and let $P_2=\cup_{i\in A_2}S_i$ be the points in $A_2$. 

Since $Z$ is $\left(\frac{1}{t},\eps\right)$-approximation for $(X,\mathcal{R})$ and $A_2$ only consists of indices $i$ such that $S_i$ contributes at least $\frac{1}{t}$ fraction of the mass, then
\begin{align}
\left(1-\eps\right)\frac{\sum_{p\in P_1}w(p)f(p,X)}{\sum_{p\in P}s(p)}\le\frac{\sum_{p\in P_2}u(p)f(p,X)}{\sum_{p\in P}s(p)}\le\left(1+\eps\right)\frac{\sum_{p\in P_1}w(p)f(p,X)}{\sum_{p\in P}s(p)}.\label{eqn:p2:approx}
\end{align}
Now if $\sum_{i=\in A_1}\mu_i\ge\frac{1}{t}$, then since $Z$ is a $\left(\frac{1}{t},\eps\right)$-approximation for $(X,\mathcal{R})$, then 
\begin{align*}
\left(1-\eps\right)\frac{\sum_{p\in P_1}w(p)f(p,X)}{\sum_{p\in P}s(p)}\le\frac{\sum_{p\in P_1}u(p)f(p,X)}{\sum_{p\in P}s(p)}\le\left(1+\eps\right)\frac{\sum_{p\in P_1}w(p)f(p,X)}{\sum_{p\in P}s(p)}.
\end{align*}
Hence combining with (\ref{eqn:p2:approx}) and noting that $P=P_1\cup P_2$, then we have
\begin{align*}
\left(1-\eps\right)\sum_{p\in P}w(p)f(p,X)\le\sum_{p\in P}u(p)f(p,X)\le\left(1+\eps\right)\sum_{p\in P}w(p)f(p,X).
\end{align*}
On the other hand, if $\sum_{i=\in A_1}\mu_i<\frac{1}{t}$, then since $Z$ is a $\left(\frac{1}{t},\eps\right)$-approximation for $(X,\mathcal{R})$, it follows that
\begin{align*}
\frac{\sum_{p\in P_1}w(p)f(p,X)}{\sum_{p\in P}s(p)}-\frac{\eps}{t}\le\frac{\sum_{p\in P_1}u(p)f(p,X)}{\sum_{p\in P}s(p)}\le\frac{\sum_{p\in P_1}w(p)f(p,X)}{\sum_{p\in P}s(p)}+\frac{\eps}{t}.
\end{align*}
Since we have $t\sum_{p\in P}m(p)\ge\sum_{p\in P}s(p)=\sum_{p\in P}\sup_{C\in Q(P)}w(p)|f(p,C)|\ge0$, then $\frac{\eps}{t}\le\frac{\eps}{\sum_{p\in P}s(p)}$ and thus
\begin{align*}
\sum_{p\in P_1}w(p)f(p,X)-\eps\le\sum_{p\in P_1}u(p)f(p,X)\le\sum_{p\in P_1}w(p)f(p,X)+\eps.
\end{align*}
Moreover, $\sum_{i=\in A_1}\mu_i<\frac{1}{t}$ implies that $\frac{\sum_{p\in S_i}w(p)f(p,X)}{\sum_{p\in P}s(p)}<\frac{\eps}{t}\le\frac{\eps}{\sum_{p\in P}s(p)}$. 
Therefore, we have $\sum_{p\in S_i}w(p)f(p,X)<1$, so that
\begin{align*}
\left(1-\eps\right)\sum_{p\in P_1}w(p)f(p,X)\le\sum_{p\in P_1}u(p)f(p,X)\le\left(1+\eps\right)\sum_{p\in P_1}w(p)f(p,X).
\end{align*}
Again combining with (\ref{eqn:p2:approx}) and noting that $P=P_1\cup P_2$, then 
\begin{align*}
\left(1-\eps\right)\sum_{p\in P}w(p)f(p,X)\le\sum_{p\in P}u(p)f(p,X)\le\left(1+\eps\right)\sum_{p\in P}w(p)f(p,X).
\end{align*}
Note that $P_2$ contains at most $\log t$ sets $S_i$, it suffices to obtain $\left(\frac{1}{t},\eps\right)$-sample for $O(\log t)$ sets, each with failure probability $\Theta\left(\frac{\delta}{\log t}\right)$, using Theorem~\ref{thm:lls:sample}.  
By a union bound, the total failure probability is at most $\delta$. 
\end{proof}

\paragraph{Intuition for Sample Complexity of Learning and Coresets.} 
The intuition is that the class of functions $\mathcal{F}$ represents the objective in the query space, so that a particular $f\in\mathcal{F}$ represents the objective for a particular query in the query space. 
For objectives like $k$-means or $k$-median clustering, each $f$ represents the objective for a separate set of $k$ centers. 
Then the goal is to learn $\mathcal{F}$ simultaneously with a small number of samples. 

The domain $X$ for the class of functions $\mathcal{F}$ translates exactly to the ground set $X$, which is the input points for objectives like $k$-means or $k$-medians. 
We first note that sampling a point of $X$ and then rescaling by the (inverse of the) sampling probability provides an unbiased estimator to the objective. 
\cite{LiLS01} then states that if we sample uniformly over $X$, we can obtain a $(1+\eps)$-approximation to the objective by bounding the variance through a small number of samples. 
Then the idea of sensitivity sampling is that instead of uniformly sampling points from $X$, we sample each point of $X$ according to its sensitivity, but still rescale by the (inverse of the) sampling probability. 
Now the expectation of the samples is still the objective, but the variance is much smaller and so we require a smaller number of samples. 

The real workhorse in this bound is Lemma~\ref{lls:lem:six} by \cite{LiLS01}, which uses the chaining technique of Kolmogorov and refined by Talagrand~\cite{talagrand1994sharper}. 
Crucially, the usage of chaining by \cite{LiLS01} manages to simultaneously learn a large number of functions in a class $\mathcal{F}$ without needing to union bound over a net over the functions in $\mathcal{F}$. 
It is precisely this technique that avoids a quadratic dependency on $t$ from the union bound. 
\section{Applications}
\label{sec:applications}
Our theoretical worst-case guarantee has a wide range of applications due to the prevalence of the coreset technique and how well-studied the total sensitivity is of various optimization problems. 
Note that for any problem whose sensitivity is known to be $t$, Theorem~\ref{thm:main} gives an improvement on dependency of $t$ from $O(t^2)$ to $O(t\log t)$ for the number of sampled points. 
\cite{VaradarajanX12b} considers sensitivity sampling for shape fitting problems, focusing on $(j,k)$-projective clustering problems, such as $k$-median/$k$-means, $k$-line clustering, $j$-subspace approximation, and the integer $(j,k)$-projective clustering problem. 
We show that our results imply more efficient coreset constructions using the total sensitivity bounds on these problems obtained by \cite{VaradarajanX12b}.

\begin{definition}[$(j,k)$-projective clustering problem]
Given integers $j,k\ge0$, a distance function $\dist:\mathbb{R}^d\times\mathbb{R}^d\to\mathbb{R}$, and a set of points $P\subseteq\mathbb{R}^d$, the goal is to find a shape $F$ in the family of shapes $\mathcal{F}$ that minimizes $\sum_{p\in P}\dist(p,F)$, where the family of shapes $\mathcal{F}$ is the set of $k$-tuples of affine $j$-subspaces. 
Hence, $F$ is the union of $k$ $j$-flats. 
\end{definition}
In particular if $j=0$, then each $j$-flat is just a point, so the $(j,k)$-projective clustering problem with the distance function $\dist(\cdot,\cdot)$ being the Euclidean distance reduces to the $k$-median problem, while if $\dist(\cdot,\cdot)$ is the squared Euclidean distance, then the objective becomes the $k$-means problem. 
If $j=1$, then each $j$-flat becomes a line, so that the objective becomes the $k$-line clustering problem. 
On the other hand, if $j$ is fixed and $k=1$, the $(j,k)$-projective clustering problem becomes the subspace approximation problem, which can be parametrized by the distance function. 
Finally, in the \emph{integer} $(j,k)$-projective clustering problem, all points in $P$ are assumed to have integer coordinates from the range $[-n^\gamma,n^\gamma]$ for some constant $\gamma>0$. 

For the $k$-line center problem and the integer general $(j,k$)-projective clustering problem, \cite{VaradarajanX12a} showed the following upper bounds on the total sensitivity:
\begin{theorem}
\cite{VaradarajanX12a}
\label{thm:vx12a:kline}
Let the distance function $\dist$ used in projective clustering be the $z$-th power of the Euclidean distance for some fixed $z\in(0,\infty)$. 
Then the total sensitivity $t$ satisfies $t=O(k^{f(d,k)}\log n)$ for $j=1$, i.e., the $k$-line center problem and $t=O((\log n)^{g(d,j,k)})$ for the integer $(j,k)$-projective clustering problem, where $f$ and $g$ are fixed functions that only depend on $d,k$ and $d,j,k$, respectively. 
\end{theorem}
Then Theorem~\ref{thm:vx12a:kline} and Theorem~\ref{thm:main} together imply efficient coresets for both the $k$-line center problem and the integer $(j,k)$-projective clustering problem.
\begin{theorem}
\label{thm:int:projective}
There exists an algorithm that outputs a set of $\tilde{O}\left(\frac{d}{\eps^2}g(d,j,k)(\log n)^{g(d,j,k)}\right)$ weighted points that is an $\epsilon$-coreset for the integer $(j,k)$-projective clustering problem with probability at least $\frac{2}{3}$. 
\end{theorem}
\begin{theorem}
\label{thm:weak:kline}
There exists an algorithm that outputs an $\epsilon$-coreset of $\tilde{O}\left(\frac{d}{\eps^2}f(d,k)k^{f(d,k)}\log n\right)$ weighted points for the $k$-line center problem with probability at least $\frac{2}{3}$. 
\end{theorem}

\noindent
We remark that Theorem~\ref{thm:weak:kline} is subsumed by Theorem~\ref{thm:tight:kline} below, as \cite{VaradarajanX12b} tighten the total sensitivity upper bound for the $k$-line center problem by showing that $f(d,k)$ in Theorem~\ref{thm:vx12a:kline} is independent of $d$. 
\begin{theorem}
\cite{VaradarajanX12b}
\label{thm:vx12b:kline}
Let the distance function $\dist$ used in projective clustering be the $z$-th power of the Euclidean distance for some fixed $z\in(0,\infty)$. 
Then the total sensitivity $t$ satisfies $t=O(k^{f(k)}\log n)$ for $j=1$, where $f$ is a fixed function that only depends on $k$, respectively. 
\end{theorem}
From Theorem~\ref{thm:vx12b:kline} and Theorem~\ref{thm:main}, we have
\begin{theorem}
\label{thm:tight:kline}
There exists an algorithm that outputs a set of $\tilde{O}\left(\frac{d}{\eps^2}f(k)k^{f(k)}\log n\right)$ weighted points that is an $\epsilon$-coreset for the $k$-line center problem with probability at least $\frac{2}{3}$. 
\end{theorem}

\noindent
\cite{VaradarajanX12b} bounded the total sensitivity for the $(0,k$)-projective clustering problem, which includes $k$-median and $k$-means. 
\begin{theorem}
\cite{VaradarajanX12b}
\label{thm:vx12b:cluster}
Let the distance function $\dist$ used in projective clustering be the $z$-th power of the Euclidean distance for some fixed $z\in(0,\infty)$. 
For $z\ge 1$, the total sensitivity $t$ satisfies $t\le 2^{2z-1}k+2^{z-1}$, whereas for $z\in(0,1)$, we have $t\le 2k+1$.  
\end{theorem}
Thus by Theorem~\ref{thm:vx12b:cluster} and Theorem~\ref{thm:main}:
\begin{theorem}
\label{thm:cluster}
For either the $k$-median problem or the $k$-means problem, there exists an algorithm that outputs a set of $O\left(\frac{d}{\eps^2}k\log k\right)$ weighted points that is an $\epsilon$-coreset, with probability at least $\frac{2}{3}$. 
\end{theorem}

\noindent
\cite{VaradarajanX12b} also bounded the total sensitivity for the $j$-subspace fitting problem. 
\begin{theorem}
\cite{VaradarajanX12b}
\label{thm:vx12b:fitting}
Let $s=\min(j,d)$ and let the distance function $\dist$ used in projective clustering be the $z$-th power of the Euclidean distance for some fixed $z\in(0,\infty)$. 
Then total sensitivity of any set of $n$ points in $\mathbb{R}^d$ for the $j$-subspace fitting problem is $O\left(s^{1+\frac{z}{2}}\right)$ for $1\le z<2$, $O(s)$ for $z=2$, and $O(s^z)$ for $z>2$. 
\end{theorem}
By Theorem~\ref{thm:vx12b:fitting} and Theorem~\ref{thm:main}, we conclude that
\begin{theorem}
\label{thm:fitting}
Let $s=\min(j,d)$ and let the distance function $\dist$ used in projective clustering be the $z$-th power of the Euclidean distance for some fixed $z\in(0,\infty)$. 
Let $t=O\left(s^{1+\frac{z}{2}}\right)$ for $1\le z<2$, $O(s)$ for $z=2$, and $O(s^z)$ for $z>2$. 
Then there exists an algorithm that outputs a set of $\tilde{O}\left(\frac{dt}{\eps^2}\right)$ weighted points that is an $\epsilon$-coreset for the $j$-subspace fitting problem with probability at least $\frac{2}{3}$. 
\end{theorem}
\newcommand{\zz}{z}
\def\c{X}
\newcommand{\gr}{X}
\newcommand{\xx}{Q}
\newcommand{\cc}{a}
\newcommand{\dd}{b}
\newcommand{\co}{C}
\newcommand{\ef}{g}
\newcommand{\pd}{g}
\newcommand{\grr}{grr}
\section{Coreset for $k$-clustering}
This section considers tighter bounds for $k$-clustering and may be skipped for general applications. 
We introduce $\rho$-pseudo distances and define the importance of a point as a generalization of the sensitivity. 
Using the importance, we then give an analog of Theorem~\ref{thm:main} with sharper bounds for $\rho$-pseudo distances. 
As a result, we obtain stronger bounds for coreset constructions for $k$-clustering. 

We reiterate that our results in this section mirror those of Section~\ref{sec:sample}. 
We only provide theoretical guarantees on the number of samples required by the sensitivity sampling framework for $k$-clustering. 
We specify the framework in full in Algorithm~\ref{one}.

We first require the following definition of $(\alpha,\beta)$-assignment for bicriteria algorithms. 
\begin{definition}[$(\alpha,\beta)$-assignment.\label{alphabeta}]
Let $\gr$ be a ground set and $(P,w,\CC,\pd)$ be a query space where $\pd:\gr^2\to[0,\infty)$. 
Let $\xx^*\in \CC(P)$ be a query that minimizes the loss of $P$ over every query,
\[
\mathrm{opt}(P)=\arg\inf_{\xx\in \CC(P)}\sum_{p\in P}w(p)\pd(p,\xx)=\sum_{p\in P}w(p)\pd(p,\xx^*).
\]
Let $\alpha,\beta> 0$, and $B\subseteq\gr$ such that $|B|\leq \beta |\xx^*|$.
A function $\mathcal{B}:P\to B$ is an $(\alpha,\beta)$-assignment for $(P,w,\CC,\pd)$ if
\[
\sum_{p\in P}w(p)\pd(p,\mathcal{B}(p))
\leq \alpha \cdot \mathrm{opt}(P).
\]
Every $b\in B$ is called a \emph{center} and its \emph{cluster in $P$} is $\mathcal{B}^{-1}(b)=\br{p\in P\mid \mathcal{B}(p)=b}$.
\end{definition}
Intuitively, an $(\alpha,\beta)$-assignment is just a bicriteria clustering with approximation factor $\alpha$ and an overselection of centers by a factor of $\beta$. 
Thus for $k$-means clustering, any $\alpha$-approximation algorithm that chooses $\beta k$ centers can be used to determine an $(\alpha,\beta)$-assignment for each of the points. 

The following definition is especially useful for $k$-means clustering. 
\begin{definition}[$\rho$-pseudo distance]\label{weakt}
Let $\gr$ be a ground set and $\rho\geq 1$. 
A symmetric function $\pd:\gr^2\to [0,\infty)$ is \emph{$\rho$-pseudo distance} over $\gr$ if for every $(p,q,\x)\in \gr^3$
\[
\pd(p,\x)\leq \rho(\pd(p,q)+  \pd(q,\x)).
\]
For a finite set $\xx\subseteq \gr$, we denote $\pd(p,\xx):=\min_{\x\in \xx}g(p,\x)$.
\end{definition}
Note that the above definition does not assume that $\pd(p,p)=0$ for every $p\in P$.
The inequality in Definition~\ref{weakt} is sometimes called ``weak triangle inequality".

\begin{lemma}\label{lemrho}
Let $\pd$ be a $\rho$-pseudo distance over a ground set $\gr$. 
For every pair of points $p,q\in \gr$ and a finite set $\xx\subseteq \gr$,
\[
\pd(p,\xx)\leq \rho(\pd(q,p)+\pd(p,\xx)).
\]
\end{lemma}
\begin{proof}
For every $p,q\in \gr$, and a center $\x_p\in \xx$ that is closest to $p$, i.e. $\pd(p,\xx)=\pd(p,\x_p)$, we have
\begin{align*}
\pd(q,\xx)=\min_{\x\in \xx}\pd(q,\x)\leq \pd(q,\x_p)\leq \rho(\pd(q,p)+\pd(p,\x_p))=\rho(\pd(q,p)+\pd(p,\xx)).
\end{align*}
\end{proof}

\newcommand{\coresetc}{\textsc{Smaller-Coreset}}
\newcommand{\coresetcond}{\textsc{Conditional-Coreset}}
\newcommand{\coresetb}{\textsc{Median-Coreset}}
\newcommand{\coresetd}{\textsc{Positive-Coreset}}
\newcommand{\s}{s}

\begin{algorithm}[!htb]
\caption{$\coreset(P,w,\mathcal{B},\s)$; see Theorem~\ref{mainthm}\label{one}}
\textbf{Input:} A weighted set $(P,w)$ where $w:P\to [0,\infty)$, an $(\alpha,\beta)$-assignment $\mathcal{B}:P\to B$ for a query space $(P,w,\CC,\pd)$, and sample size (integer) $\s\geq 1$.\\
\textbf{Output:} A weighted set $(C,u)$.
\For{every center $b\in B$ and a point in its cluster $p\in \mathcal{B}^{-1}(b)$}
{
Set $\displaystyle \pr(p):=  \frac{w(p)\pd(p,\mathcal{B}(p))}{2\sum_{q\in P}w(q)\pd(q,\mathcal{B}(q))}
+\frac{w(p)}{2|B|\sum_{q\in \mathcal{B}^{-1}(b)}w(q)}$.\label{l4}\\
\tcp{Note that \quad $\sum_{p\in P}\pr(p)=1$}}
$C:=\emptyset$\\
\For{$\s$ iterations}{
Sample a point $q$ from $P$, where every $p\in P$ is chosen with probability $\pr(p)$.\label{l5algcore}\\
$C:= C\cup\br{q}$\tcp{add $q$ to the multi-set $C$}\label{l6algcore}
$\displaystyle u(q):=\frac{w(q)}{|C|\cdot \pr(q)}
$}
\Return $(C,u)$
\end{algorithm}
We now give a generalization of the notion of sensitivity in the form of importance. 
\newcommand{\yy}{f}
\begin{definition}[The function $f$ and its importance $m$\label{ff}]
Let $(P,w,\CC,\pd)$ be a query space where $w:P\to[0,\infty)$ and $\pd$ is a $\rho$-pseudo distance.
Let $\mathcal{B}:P\to B$ be an $(\alpha,\beta)$-assignment for $(P,w,\CC,\pd)$. For every $p\in P$ and $\xx\in \CC(P)$, we define
\begin{equation}\label{ffdef}
\yy(p,\xx)=\frac{\pd(p,\xx)}{\sum_{q\in P}w(q)\pd(q,\xx)}.
\end{equation}
For every center $b\in B$ and a point $p\in \mathcal{B}^{-1}(b)$ in its cluster we define
\[
m(p)=\frac{\rho\alpha \cdot w(p)\pd(p,\mathcal{B}(p))}{\sum_{q\in P}w(q)\pd(q,\mathcal{B}(p))}
+\frac{\rho^2(\alpha+1)w(p)}{\sum_{q\in \mathcal{B}^{-1}(b)}w(q)}.
\]
\end{definition}

We now show that the importance $m$ satisfies a similar function as the notion of sensitivity. 
\begin{lemma}\label{sensbound}
Let $\mathcal{B}:P\to B$, $m$ and $\yy$ be defined as in Definition~\ref{ff}.
Then for every $p\in P$ we have $$m(p)\geq \sup_{\xx\in \CC(P)} w(p)\yy(p,\xx).$$
\end{lemma}
\begin{proof}
For simplicity, we denote $p'=\mathcal{B}(p)$ and $q'=\mathcal{B}(q)$ for every $p,q\in P$, and $P_b=\mathcal{B}^{-1}(p)$ for every $b\in B$.
We prove the claim for a point $p\in P_b$ in the cluster of some center $b\in B$, as in Definition~\ref{ff}. Let $\xx\in \CC(P)$ and assume $w(p)\yy(p,\xx)>0$, otherwise the lemma trivially holds.
We need to upper bound
\begin{align}
w(p)f(p,\xx)&=\frac{w(p)\pd(p,\xx)}{\sum_{q\in P}w(q)\pd(q,\xx)}\leq \frac{\rho w(p) \pd(p,p')}{\sum_{q\in P}w(q)\pd(q,\xx)}+\frac{\rho w(p)\pd(p',\xx)}{\sum_{q\in P}w(q)\pd(q,\xx)}\nonumber\\
&\leq \frac{\alpha\rho w(p)\pd(p,p')}{\sum_{q\in P}w(q)\pd(q,q')}+\frac{\rho w(p)\pd(p',\xx)}{\sum_{q\in P}w(q)\pd(q,\xx)},\label{aa}
\end{align}
where the first inequality holds by Lemma~\ref{lemrho}, and the second inequality holds since $\mathcal{B}$ is an $(\alpha,\beta)$-assignment.
To bound the last term, note that
\begin{align}
\pd(p',\xx)\sum_{q\in P_b}w(q)&=\sum_{q\in P_b}w(q)\pd(p',\xx)\leq \sum_{q\in P_b}w(q)\cdot \rho( \pd(p',q)+\pd(q,\xx))\nonumber\\
&= \rho\sum_{q\in P_b}w(q) \pd(q,b)+\rho\sum_{q\in P_b}w(q) \pd(q,\xx)\label{eqeq}\\
&\leq \rho\alpha\sum_{q\in P_b}w(q) \pd(q,\xx)+\rho\sum_{q\in P_b}w(q) \pd(q,\xx) = \rho(\alpha+1)\sum_{q\in P_b}w(q) \pd(q,\xx),\label{eq161}
\end{align}
where the first inequality is by Lemma~\ref{lemrho},~\eqref{eqeq} holds since $p'=b$ and since $g$ is symmetric by definition, and~\eqref{eq161} holds since $\mathcal{B}$ is an $(\alpha,\beta)$-assignment.

Dividing by $\sum_{q\in P_b}w(q)\cdot \sum_{q\in P}w(q)\pd(q,\xx)$ yields
\[
\frac{\pd(p',\xx)}{\sum_{q\in P}w(q)\pd(q,\xx)} \leq \frac{\rho(\alpha+1)}{\sum_{q\in P_b}w(q)}.
\]
Substituting this in~\eqref{aa} yields the desired result
\begin{align*}
w(p)f(p,\xx)&\leq \frac{\rho\alpha w(p)\pd(p,p')}{\sum_{q\in P}w(q)\pd(q,q')}+\frac{\rho^2(\alpha+1)w(p)}{\sum_{q\in P_b}w(q)}\\
&= m(p).
\end{align*}
\end{proof}

As a warm-up, we now prove an analog of Theorem~\ref{thm:main} for $\rho$-pseudo distances that provides tighter bounds, due to the tighter setting of $t$ from the $(\alpha,\beta)$ assignment.
\begin{theorem}
\label{mainthm}
Let
\begin{itemize}
\item $(P,w,\CC,\pd)$ be a query space, where $\pd$ is a $\rho$-pseudo distance and $w:P\to[0,\infty)$.
\item $\mathcal{B}:P\to B$ be an $(\alpha,\beta)$ assignment for $(P,w,\CC,\pd)$. 
\item $d$ be the VC-dimension of $(P,w,\CC,\yy)$, where $f$ was defined in~\eqref{ffdef}.
\item $t=2(\rho\alpha+\rho^2(\alpha+1)|B|)$.
\item $c\geq1$ be a sufficiently large constant, $\eps,\delta\in(0,1)$, and
\[
 \s\geq\frac{ct}{\eps^2}\left(d\log t+\log\left(\frac{1}{\delta}\right)  \right).
\]
\item $(\co,u)$ be the output of a call to $\coreset(P,w,\mathcal{B},\s)$; see Algorithm~\ref{one}.
\end{itemize}
Then, $\co\subseteq P$, $u:\co\to[0,\infty)$ and, with probability at least $1-\delta$, $(C,u)$ is an $\eps$-coreset of size $|C|=\s$ for $(P,w,\CC,\pd)$.
\end{theorem}
\begin{proof}
Let $p\in P$. By Lemma~\ref{sensbound}, for every $\xx\in \CC(P)$
\begin{equation}\label{mpp}
\m(p)\geq w(p)f(p,\xx).
\end{equation}
The probability of choosing $p$ to be, say, the first point in $\co$, is
\begin{align*}
\pr(p)&=\frac{w(p)\pd(p,\mathcal{B}(p))}{2\sum_{q\in P}w(q)\pd(q,\mathcal{B}(q))}+
\frac{w(p)}{2|B|\sum_{q\in  \mathcal{B}^{-1}(b)}w(q)}\\
&\geq \frac{1}{2(\rho\alpha+\rho^2(\alpha+1)|B|)}\cdot\Bigg(\frac{\rho\alpha w(p)\pd
(p,\mathcal{B}(p))}{\sum_{q\in P}w(q)\pd(q,\mathcal{B}(q))}+\frac{\rho^2(\alpha+1)w(p)}{\sum_{q\in  \mathcal{B}^{-1}(b)}w(q)}\Bigg)\geq \frac{m(p)}{t}.
\end{align*}
Using the last inequality and~\eqref{mpp}, we apply Theorem~\ref{insofcor} to obtain that, with probability at least $1-\delta$, we have that for all $\xx\in \CC(C)$,
\[
\left|\sum_{p\in P} w(p)f(p,\xx)-\sum_{q\in C} \frac{w(q)}{\pr(q)|C|}\cdot f(q,\xx)\right| \leq \eps.
\]
Multiplying this by $\sum_{q\in P}w(q)\pd(q,\xx)$, and substituting $u(q)=\frac{w(q)}{\pr(q)|C|}$ yields that for all $\xx\in \CC(C)$,
\[
\left|\sum_{p\in P} w(p)g(p,\xx)-\sum_{q\in C} u(q)\cdot g(q,\xx)\right| \leq \eps\sum_{q\in P}w(q)\pd(q,\xx)
\]
implies that $(C,u)$ is an $\eps$-coreset as desired.
\end{proof}
\newcommand{\argmin}{\mathrm{argmin}}

\subsection{Smaller coreset}
In this section, we define a $(\rho,\psi,\phi)$-pseudo distance function, which serves as a generalization of $\rho$-pseudo distances, and give smaller coreset constructions for $(\rho,\psi,\phi)$-pseudo distance functions. 
Our algorithms appear in Algorithm~\ref{algmed} and Algorithm~\ref{algmed5}. 
\begin{algorithm}[!htb]
\caption{$\coresetc(P,w,\mathcal{B},s)$; See Theorem~\ref{mainthm2}\label{algmed}}
\textbf{Input:} \quad\quad\= A weighted set $(P,w)$ where $w:P\to[0,\infty)$, $(\alpha,\beta)$-assignment $\mathcal{B}:P\to B$ for $(P,w,\CC,\pd)$, and sample size (integer) $s\geq 1$.\\
\textbf{Output:} A weighted set $(C\cup B,u)$.\\
$(C,u):= \coreset(P,w,\mathcal{B},s)\quad$\tcp{see Algorithm~\ref{one}}
\For{every $b\in B$}{
{Set $u(b):=\sum_{p\in \mathcal{B}^{-1}(b)}w(p)-\sum_{q\in C\cap \mathcal{B}^{-1}(b)}u(q)$\label{seven}}}
\Return $(C\cup B,u)$\tcp{\quad$C\cup B$ is a multi-set}
\end{algorithm}

We define the following generalization of distance to handle $k$-clustering. 
\begin{definition}\label{ass1}
Let $g$ be a $\rho$-pseudo distance over $\gr$ as in Definition~\ref{weakt}. 
For $\phi>0$ and $\psi\geq 0$, $g$ is also a \emph{$(\rho,\psi,\phi)$-pseudo distance function} if for every $(p,q,\x)\in \gr^3$ we have
\[
|\pd(p,\x)-\pd(q,\x)|\leq \phi \pd(p,q)+ \psi \pd(q,\x).
\]
\end{definition}
Intuitively, the $\rho$-pseudo distance handles loss functions such as the squared Euclidean distance that do not satisfy the triangle inequality but rather a generalized version of the triangle inequality. 

\begin{lemma}\label{eqq}
Let $\pd:\gr^2\to[0,\infty)$ be a $(\rho,\psi,\phi)$-pseudo distance function. Then for every finite set $M\subseteq \gr$ and $p,q\in M$ we have
\[
|\pd(p,\xx)-\pd(q,\xx)|\leq \phi \pd(p,q)+ \psi \pd(q,\xx).
\]
\end{lemma}
\begin{proof}
We assume that $M$ is non-empty, otherwise the claim trivially holds, as we define the minimum of an empty set to be $0$ in the notation. Let $\x_q\in\textrm{argmin}_{\x\in \xx}\pd(q,x)$ and $\x_p\in\argmin_{x\in \xx}\pd(p,x)$. The proof is by case analysis: (i) $\pd(p,\xx)>\pd(q,\xx)$, and (ii) $\pd(p,\xx)\leq \pd(q,\xx)$ as follows.\\
\textbf{Case (i): $\pd(p,\xx)>\pd(q,\xx)$.} We have
\begin{align}
|\pd(p,\xx)-\pd(q,\xx)|&=\pd(p,\xx)-\pd(q,\xx)=\pd(p,\xx)-\pd(q,\x_q)\nonumber\\
&\leq \pd(p,\x_q)-\pd(q,\x_q)\leq \phi \pd(p,q)+ \psi \pd(q,\x_q)\nonumber\\
&= \phi \pd(p,q)+ \psi \pd(q,\xx),\label{tt1}
\end{align}
where the first inequality is by the definition of $\pd(p,\xx)=\min_{\x\in \xx}\pd(p,\xx)$, and the second inequality is by Definition~\ref{ass1}.

\noindent\textbf{Case (ii): $\pd(p,\xx)\leq \pd(q,\xx)$. }
We have
\begin{align}
|\pd(p,\xx)-\pd(q,\xx)|&=\pd(q,\xx)-\pd(p,\xx) =\pd(q,\x_q)-\pd(p,\x_p)\nonumber\\
&\leq \pd(q,\x_p)-\pd(p,\x_p)\leq \phi \pd(q,p)+\psi \pd(p,\x_p)\nonumber\\
&= \phi \pd(q,p)+\psi \pd(p,\xx)\leq \phi \pd(p,q)+\psi \pd(q,\xx),\label{tt2}
\end{align}
where the last inequality is by the assumption of this case. Combining~\eqref{tt1} and~\eqref{tt2} yields that (in both cases)
\[
|\pd(p,\xx)-\pd(q,\xx)|\leq \phi \pd(p,q)+ \psi \pd(q,\xx).
\]
\end{proof}

\newcommand{\hh}{h}
\begin{definition}[Conditional normalized distance $\hh$.]\label{assnew}
Let $(P,w,\CC,\pd)$ be a query space where $w:P\to[0,\infty)$, and $\pd:\gr^2\to[0,\infty)$ is a $(\rho,\psi,\phi)$-pseudo distance function. 
Let $\mathcal{B}:P\to B$ be an $(\alpha,\beta)$-assignment for $(P,w,\CC,\pd)$.
Let $\eps\in (0,1)$ such that $\psi<\eps/(4\rho(\alpha+1))$, and
\begin{equation}\label{epss}
\eps'=\frac{\eps}{4\phi\rho(\alpha+1)}-\frac{\psi}{\phi}.
\end{equation}
For every $\xx\in \CC(P)$ define
\begin{equation}\label{Fdef}
F(\xx)=\br{p\in P\mid \pd(\mathcal{B}(p),\xx)>\frac{\pd(p,\mathcal{B}(p))}{\eps'}},
\end{equation}
and $\hh:P\times \CC(P)\to \REAL$ such that for every $p\in P$ and $\xx\in \CC(P)$,
\begin{equation}\label{hhh}
\hh(p,\xx)=
\begin{cases}
\frac{\pd(p,\xx)-\pd(\mathcal{B}(p),\xx)}{\sum_{q\in P}w(q)\pd(q,\xx)} & p\in P\setminus F(\xx)\\
0 & p\in F(\xx).
\end{cases}
\end{equation}
\end{definition}

\begin{theorem}\label{mainthm2}
Consider the variables in Definition~\ref{assnew}.
Let
\[
t=1+2\alpha(\phi +\psi/\eps')
\]
and $d$ be the VC-dimension of $(P,w,\CC,\hh)$.
Let $c'$ be a sufficiently large constant,
\[
  s\geq\frac{c't}{\eps^2}\left(d\log t+\log\left(\frac{1}{\delta}\right)  \right)
  +c'|B|\left(\log |B|+\log\left(\frac{1}{\delta}\right) \right),
\]
and $(C\cup B,u)$ be the output of a call to algorithm $\coresetc(P,w,\mathcal{B},s)$; see Algorithm~\ref{algmed}.
Then, $C\subseteq P$, $u:C\to[0,\infty)$, and with probability at least $1-\delta$, we have that for all $X\in\CC(C)$,
\[
\left|\sum_{p\in P}w(p)\pd(p,X)   -\sum_{q\in C\cup B}u(q)\pd(q,X)\right|\leq \eps\sum_{p\in P}w(p)\pd(p,X).
\]
\end{theorem}
\begin{proof}
Let $\xx\in \CC(C)$ and extend the function $u$ as defined in Algorithm~\ref{algmed} to be $u(p)=0$ for every $p\in P\setminus C$. Also define $v(p)=w(p)-u(p)$ and $p'=\mathcal{B}(p)$ for every $p\in P$.
The difference in the loss between taking the original points or its coreset $C\cup B$ is
\begin{align}
|\sum_{p\in P}w(p)\pd(p,\xx)-&\sum_{q\in C\cup B}u(q)\pd(q,\xx)|=|\sum_{p\in P}v(p)\pd(p,\xx)-\sum_{b\in B}u(b)\pd(b,\xx)|\nonumber\\
&\leq \left|\sum_{p\in P}v(p)(\pd(p,\xx)-\pd(p',\xx))\right|+\left|\sum_{p\in  P}v(p)\pd(p',\xx)-\sum_{b\in B}u(b)\pd(b,\xx)\right|\label{toprove33}\\
&=\left|\sum_{p\in P}v(p)(\pd(p,\xx)-\pd(p',\xx))\right|,\label{last4}
\end{align}
where~\eqref{toprove33} is by the triangle inequality, and~\eqref{last4} holds since
\begin{align*}
\sum_{p\in  P}v(p)\pd(p',\xx)&=\sum_{b\in  B}\sum_{p\in\mathcal{B}^{-1}(b)}v(p)\pd(b,\xx)=\sum_{b\in B}\pd(b,\xx)\sum_{p\in \mathcal{B}^{-1}(b)}(w(p)-u(p))\\
&=\sum_{b\in B}\pd(b,\xx)\Bigg(\sum_{p\in \mathcal{B}^{-1}(b)}w(p)-\sum_{q\in C\cap \mathcal{B}^{-1}(b)}u(q)\Bigg)\\
&=\sum_{b\in B}u(b)\pd(b,\xx),
\end{align*}
where the last equality is by the definition of $u$ in Line~\ref{seven} of Algorithm~\ref{algmed}.

By letting $F=F(\xx)$ as defined in~\eqref{Fdef},~\eqref{last4} is bounded by $\left|\sum_{p\in P}v(p)(\pd(p,\xx)-\pd(p',\xx))\right|$, which equals
\begin{align}
\Bigg|\sum_{p\in P\setminus F}v(p)&(\pd(p,\xx)-\pd(p',\xx))+\sum_{p\in F}v(p)(\pd(p,\xx)-\pd(p',\xx))\Bigg|\nonumber\\
&\leq \left|\sum_{p\in P\setminus F}v(p)(\pd(p,\xx)-\pd(p',\xx))\right|\label{toprove31}\\
&+\left|\sum_{p\in F}v(p)(\pd(p,\xx)-\pd(p',\xx))\right|\label{toprove32},
\end{align}
where the last inequality is by the triangle inequality. 
We now bound each of the last terms.

\textbf{Bound on~\eqref{toprove31}}:
Let $\hh:P\times \CC(M)\to\REAL$ be as in Definition~\ref{assnew}. 
Let $p\in P\setminus F$ and recall that $\pr(p)$ was defined to be the probability of choosing $p$ in Line~\ref{l5algcore} of Algorithm~\ref{one}. We then have
\begin{align}
w(p)|\hh(p,\xx)|&=\frac{w(p)|\pd(p,\xx)-\pd(p',\xx)|}{\sum_{q\in P}w(q)\pd(q,\xx)}\nonumber\\
&\leq \frac{\alpha\cdot w(p)|g(p,\xx)-\pd(p',\xx)|}{\sum_{q\in P}w(q)\pd(q,q')}\label{aaa10}\\
&\leq \frac{\alpha\cdot w(p) (\phi \pd(p,p')+\psi \pd(p',\xx))}{\sum_{q\in P}w(q)\pd(q,q')}\label{aa3}\\
&\leq \frac{\alpha\cdot w(p) \pd(p,p')(\phi +\psi/\eps')}{\sum_{q\in P}w(q)\pd(q,q')}\label{aa33}\\
&\leq 2\alpha(\phi +\psi/\eps') \pr(p)\nonumber,
\end{align}
where~\eqref{aaa10} holds since $\mathcal{B}$ is an $(\alpha,\beta)$-assignment, i.e.,
\begin{equation}\label{BB}
\sum_{q\in P}w(q)g(q,q')\leq \alpha\sum_{q\in P}w(q)g(q,\xx),
\end{equation}
\eqref{aa3} is by Lemma~\ref{eqq}, and~\eqref{aa33} is by~\eqref{Fdef} and the assumption $p\in P\setminus F$.

Hence, $\sup_{\xx\in \CC(M)}w(p)|\hh(p,\xx)|$ is bounded by $\m(p)=2\alpha(\phi +\psi/\eps') \pr(p)$,
\[
\pr(p)=\frac{m(p)}{2\alpha(\phi +\psi/\eps')}\geq \frac{m(p)}{t},
\]
$t\geq 2$, and for every constant $c>0$ there is a sufficiently large constant $c'$ such that
\[
|\co|=s\geq
\frac{c't}{\eps^2}\left(d\log t+\log\left(\frac{1}{\delta}\right)  \right)
  \geq
\frac{16ct}{\eps^2}\left(d\log t+\log\left(\frac{2}{\delta}\right)\right).
\]
Plugging these bound in Theorem~\ref{insofcor} with $\eps/4$, and the query space $(P,w,\CC,\hh)$ yields that, with probability at least $1-\delta/2$, we have that for all $\xx\in \CC(C)$,
\begin{align*}
\left|\sum_{p\in P}w(p)\hh(p,\xx)-\sum_{q\in C}\frac{w(q)}{|C|\pr(q)}\cdot\hh(q,\xx)\right|\leq \frac{\eps}{4}.
\end{align*}
Substituting $u(q)=w(q)/(|C|\pr(q))$, $v(q)=w(q)-u(q)$ for every $q\in P$, and removing the points $p\in F(\xx)$ whose loss is $\hh(p,\xx)=0$ simplify the last expression to
\begin{align*}
\forall\xx\in \CC(C): \left|\sum_{p\in P\setminus F}v(p)\hh(p,\xx)\right|\leq \frac{\eps}{4}.
\end{align*}

Assume that the last equation indeed holds (which happens with probability at least $1-\delta/2$). 
By this and the definition of $g$, for every $\xx\in \CC(C)$,~\eqref{toprove31} is bounded by
\begin{align}
\Bigg|\sum_{p\in P\setminus F}v(p)(\pd(p,\xx)-\pd(p',\xx))\Bigg|&=\left|\sum_{p\in P\setminus F}v(p)\hh(p,\xx)\right|\cdot \sum_{q\in P}w(q)\pd(q,\xx)\nonumber\\
&\leq \frac{\eps}{4} \sum_{q\in P} w(q)\pd(q,\xx)\label{qqq}.
\end{align}

\textbf{Bound on~\eqref{toprove32}:} Since $|v(p)|=|w(p)-u(p)|\leq w(p)+u(p)$, and using the triangle inequality
\begin{align}
\Bigg|\sum_{p\in F}\frac{v(p)(\pd(p,\xx)-\pd(p',\xx))}{\sum_{q\in P}w(q)\pd(q,\xx)}\Bigg|&\leq \sum_{p\in F}\frac{|v(p)|\cdot |\pd(p,\xx)-\pd(p',\xx)|}{\sum_{q\in P}w(q)\pd(q,\xx)}\nonumber\\
&\leq \sum_{p\in F}\frac{(w(p)+u(p))|\pd(p,\xx)-\pd(p',\xx)|}{\sum_{q\in P}w(q)\pd(q,\xx)}\label{tri3}.
\end{align}
To bound the numerator, 
\begin{align}
|\pd(p,\xx)-\pd(p',\xx)|&\leq \phi \pd(p,p')+\psi \pd(p',\xx)\leq(\phi \eps'+\psi) \pd(p',\xx)\nonumber\\
&\leq \frac{\eps \pd(p',\xx)}{4\rho(\alpha+1)}\label{dpcc},
\end{align}
where the first inequality is by Lemma~\ref{eqq}, the second holds since $p\in F$, and the last inequality is by~\eqref{epss}.

Our $(\alpha,\beta)$-assignment approximates the sum of distances to a query up to an additive error as follows.
\begin{align}
\label{firstin2}\sum_{q\in P}w(q)\pd(q',\xx)
&\leq \sum_{q\in P}w(q) \rho(\pd(q',q)+\pd(q,\xx))\\
\nonumber&= \rho\sum_{q\in P}w(q)\pd(q',q)+\rho\sum_{q\in P}w(q)\pd(q,\xx)\\
\label{secin2}&\leq \rho(\alpha+1)\sum_{q\in P}w(q)\pd(q,\xx), 
\end{align}
where~\eqref{firstin2} is by Lemma~\ref{lemrho}, and~\eqref{secin2} is by~\eqref{BB}. 
Hence,
\begin{align}
\Bigg|\sum_{p\in F}\frac{v(p)(\pd(p,\xx)-\pd(p',\xx))}{\sum_{q\in P}w(q)\pd(q,\xx)}\Bigg|&\leq \sum_{p\in F}\frac{(w(p)+u(p))|\pd(p,\xx)-\pd(p',\xx)|}{\sum_{q\in P}w(q)\pd(q,\xx)}\label{align1}\\
&\leq \sum_{p\in F}\frac{(\rho\alpha+1)(w(p)+u(p))|\pd(p,\xx)-\pd(p',\xx)|}{\sum_{q\in P}w(q)\pd(q',\xx)}\label{align11}\\
&\leq \frac{\eps}{4}\sum_{p\in F}\frac{(w(p)+u(p))\pd(p',\xx)}{\sum_{q\in P}w(q)\pd(q',\xx)}\label{align2}\\
&\leq \frac{\eps}{4}\sum_{p\in P}\frac{w(p)\pd(p',\xx)}{\sum_{q\in P}w(q)\pd(q',\xx)}+\frac{\eps}{4} \sum_{p\in P}\frac{u(p)\pd(p',\xx)}{\sum_{q\in P}w(q)\pd(q',\xx)}\nonumber\\
&=\frac{\eps}{4}+\frac{\eps}{4} \sum_{p\in P}\frac{u(p)\pd(p',\xx)}{\sum_{q\in P}w(q)\pd(q',\xx)},\label{align4}
\end{align}
where~\eqref{align1} is by~\eqref{tri3},~\eqref{align11} holds by~\eqref{firstin2},~\eqref{align2} holds by~\eqref{dpcc}, and~\eqref{align4} holds since $F\subseteq P$.

It is left to bound the rightmost term in~\eqref{align4}. Let $\zz:P\times B\to[0,\infty)$ such that for every $b\in B$ and $p\in P_b$
\begin{equation}\label{hhh2}
\zz(p,b)=
\begin{cases}
\frac{1}{\sum_{q\in P_b}w(q)} & p\in P_b \\
0 & p\in P\setminus P_b.
\end{cases}
\end{equation}
The last term in~\eqref{align4} is then bounded by
\begin{align}
\sum_{p\in P}\frac{u(p)\pd(p',\xx)}{\sum_{q\in P}w(q)\pd(q',\xx)}&=\sum_{b\in B}\sum_{p\in P_b}\frac{u(p)\pd(p',\xx)}{\sum_{q\in P}w(q)\pd(q',\xx)}\label{sj1}\\
\label{sj2}&=\sum_{b\in B}\frac{\sum_{q\in P_b}w(q)\pd(q',\xx)}{\sum_{q\in P}w(q)\pd(q',\xx)}\sum_{p\in P_b}\frac{u(p)\pd(p',\xx)}{\sum_{q\in P_b}w(q)\pd(q',\xx)}\\
\label{sj3}&=\sum_{b\in B}\frac{\sum_{q\in P_b}w(q)\pd(q',\xx)}{\sum_{q\in P}w(q)\pd(q',\xx)}\sum_{p\in P_b}\frac{u(p)}{\sum_{q\in P_b}w(q)}\\
\nonumber&=\sum_{b\in B}\frac{\sum_{q\in P_b}w(q)\pd(q',\xx)}{\sum_{q\in P}w(q)\pd(q',\xx)}\sum_{p\in P}u(p)\zz(p,b),
\end{align}
where~\eqref{sj1} holds since $P=\bigcup_{b\in B}P_b$, in~\eqref{sj2} we simply multiplied and divided by $\sum_{q\in P_b}w(q)\pd(q',\xx)$, and~\eqref{sj3} holds since $p'=q'$ for every $p,q\in P_b$.

For every $b\in B$, we have 
\begin{align}
\sum_{p\in P}u(p)\zz(p,b)&=\sum_{p\in P}w(p)\zz(p,b)-\sum_{p\in P}v(p)\zz(p,b)\label{above44}\\
&=\sum_{p\in P_b}w(p)\zz(p,b)-\sum_{p\in P}v(p)\zz(p,b)\nonumber\\
&=1-\sum_{p\in P}v(p)\zz(p,b)\leq 1+\left|\sum_{p\in P}v(p)\zz(p,b)\right|,\label{above46}
\end{align}
where~\eqref{above44} holds since $v(p)=w(p)-u(p)$, and~\eqref{above46} is by definition~\eqref{hhh2} of $\zz$.

Let $b\in B$, $t'=2|B|$, and for every $p\in P$, let $m(p)=w(p)\zz(p,b)$. Hence, for every $p\in P$,
\begin{align*}
\pr(p) &\geq \frac{w(p)}{2|B|\sum_{q\in \mathcal{B}^{-1}(p)}w(q)} = \frac{w(p)\zz(p,\mathcal{B}(p))}{2|B|}\\
&\geq \frac{w(p)\zz(p,b)}{2|B|} = \frac{m(p)}{t'},
\end{align*}
and for every constant $c\geq1$ there is a sufficiently large $c'$ such that
\begin{align*}
|C|=s&\geq 4ct'\left(\log t'+\log\left(\frac{|B|}{\delta}\right)\right)\in O(|B|)\left(\log |B|+\log\left(\frac{2|B|}{\delta}\right)\right)
\end{align*}

Substituting the query space $(P,w,\br{b},\zz)$, $\eps=1/2$, $d=1$, and $\delta/|B|$ instead of $\delta$ in Theorem ~\ref{insofcor}, yields that with probability at least $1-\delta/(2|B|)$, we have
\begin{equation}\label{lastlast3}
\left|\sum_{p\in P}w(p)\zz(p,b)-\sum_{q\in C}u(q)\zz(q,b)\right|\leq \frac{1}{2}. 
\end{equation}
Assume the event that~\eqref{lastlast3} holds for every $b\in B$ occurs, which happens with probability at least $\delta/2$, by the union bound\footnote{Instead of using the union bound, we could simply choose $B$ as the set of queries, $\delta$ instead of $\delta/(2|B|)$ and $d=\log |B|$. However, in this would introduce a term of $d\log t=O(\log^2|B|)$ in the coreset size compared to the current $\log |B|$ term.}. Plugging~\eqref{lastlast3} in~\eqref{above46} yields
\begin{equation}\label{ssj}
\sum_{p\in P}u(p)\zz(p,b)\leq 1+\left|\sum_{p\in P}v(p)\zz(p,b)\right|\leq 2.
\end{equation}
Combining the last inequalities bounds~\eqref{toprove32} with probability at least $1-\delta/2$, as
\begin{align}
\Bigg|\sum_{p\in F}\frac{v(p)(\pd(p,\xx)-\pd(p',\xx))}{\sum_{q\in P}w(q)\pd(q,\xx)}\Bigg|&\leq \frac{\eps}{4}+\frac{\eps}{4} \sum_{p\in P}\frac{u(p)\pd(p',\xx)}{\sum_{q\in P}w(q)\pd(q',\xx)}\label{ttr1}\\
&= \frac{\eps}{4}+\frac{\eps}{4}\sum_{b\in B}\frac{\sum_{q\in P_b}w(q)\pd(q',\xx)}{\sum_{q\in P}w(q)\pd(q',\xx)}\sum_{p\in P}u(p)\zz(p,b)\label{ttr2}\\
&\leq \frac{\eps}{4}+\frac{\eps}{2}=\frac{3\eps}{4},\label{ttr3}
\end{align}
where~\eqref{ttr1} holds by~\eqref{align4},~\eqref{ttr2} by~\eqref{sj1}, and~\eqref{ttr3} by~\eqref{ssj}.

Finally, replacing~\eqref{toprove31} and~\eqref{toprove32} with~\eqref{qqq} and~\eqref{ttr3} respectively, proves that, with probability at least $1-\delta/2-\delta/2=1-\delta$ we have
\begin{align}
\Bigg|\sum_{p\in P}v(p)(\pd(p,\xx)-\pd(p',\xx))\Bigg|&\leq \left|\sum_{p\in P\setminus F}v(p)(\pd(p,\xx)-\pd(p',\xx))\right|+\left|\sum_{p\in F}v(p)(\pd(p,\xx)-\pd(p',\xx))\right|\nonumber\\
&\leq \frac{\eps}{4}\sum_{q\in P}w(q)\pd(q,\xx)+\frac{3\eps}{4}\sum_{q\in P}w(q)\pd(q,\xx)\nonumber\\
&=\eps\sum_{q\in P}w(q)\pd(q,\xx).\nonumber
\end{align}
By this and~\eqref{last4}, it follows that $(C,u)$ approximates $X$ as desired.
\end{proof}

We now handle the specific case where $\pd:\gr^2\to[0,\infty)$ is a $(\rho,\psi,\phi)$-pseudo distance function. 

\begin{theorem}\label{mainthm22}
Consider the variables in Theorem~\ref{mainthm2}, where $s$ is replaced by
\[
  s\geq\frac{c't(1+\rho(\alpha+1))^2}{\eps^2}\left(d\log t+\log\left(\frac{1}{\delta}\right)  \right)
  +\frac{c'|B|(1+\rho(\alpha+1))^2}{\eps^2}\left(\log |B|+\log\left(\frac{1}{\delta}\right)  \right).
\]
Let $(C,u)$ be the output of a call to algorithm $\coreset(P,w,\mathcal{B},\s)$; see Algorithm~\ref{algmed}.

Then, $C\subseteq P$, $u:C\to[0,\infty)$, and with probability at least $1-\delta$, $(C,u)$ is an $\eps$-coreset of size $s$ for $(P,w,\CC,\pd)$.
\end{theorem}
\begin{proof}
Let $\eps'=\eps/(1+\rho(\alpha+1))$.
After replacing $\delta$ with $\delta/2$ and $\eps$ with $\eps'$ in Theorem~\ref{mainthm2}, we obtain that with probability at least $1-\delta/2$,
\begin{equation}\label{above5}
\forall X\in \CC(C): \left|\sum_{p\in P}w(p)\pd(p,X)   -\sum_{q\in C\cup B}u(q)\pd(q,X)\right|
\leq \eps'\sum_{p\in P}w(p)\pd(p,X).
\end{equation}
Assume that this event indeed occurs and the inequality holds, and let $\xx\in \CC(C)$.

We will bound the error by excluding $B$ from this coreset, i.e.,
\begin{align}
\nonumber& \left|\sum_{p\in P}w(p)\pd(p,X)   -\sum_{q\in C}u(q)\pd(q,X)\right|\\
\label{tri1}& \leq \left|\sum_{p\in P}w(p)\pd(p,X)   -\sum_{q\in C\cup B}u(q)\pd(q,X)\right|
 +\left|\sum_{q\in C\cup B}u(q)\pd(q,X)-\sum_{p\in C}u(p)\pd(p,X) \right|\\
\label{tri2} &\leq \eps'\sum_{p\in P}w(p)\pd(p,X)+\left|\sum_{q\in C\cup B}u(q)\pd(q,X)-\sum_{p\in C}u(p)\pd(p,X) \right|,
\end{align}
where~\eqref{tri1} is by the triangle inequality, and~\eqref{tri2} is by~\eqref{above5}.
The rightmost term is
\begin{align}
\begin{split}
\left|\sum_{q\in C\cup B}u(q)\pd(q,X)-\sum_{p\in C}u(p)\pd(p,X) \right|
&=\left|\sum_{q\in B}u(q)\pd(q,X)\right|\\
&\label{sece}=\left|\sum_{b\in B}\pd(b,X)\left(\sum_{p\in P_b}w(p)-\sum_{p\in C\cap P_b}u(p) \right)\right|\\
&\leq \sum_{b\in B}\pd(b,X)\left|\sum_{p\in P_b}w(p)-\sum_{p\in C\cap P_b}u(p)\right|,
\end{split}
\end{align}
where~\eqref{sece} is by the definition of $u$ in Line~\ref{seven} of Algorithm~\ref{algmed}.

The bound on the rightmost term is similar to~\eqref{lastlast3}, after replacing the bound $1/2$ with $\eps'$, which is the reason for the largest size $s$ of the coreset. Specifically, let $b\in B$, $t'=2|B|$, and for every $p\in P$, let $m(p)=w(p)\zz(p,b)$, where $\zz$ is defined in~\eqref{hhh2}. Hence, for every $p\in P$,
\[
\pr(p) \geq \frac{w(p)}{2|B|\sum_{q\in \mathcal{B}^{-1}(p)}w(q)}
= \frac{w(p)\zz(p,\mathcal{B}(p))}{2|B|}
\geq \frac{w(p)\zz(p,b)}{2|B|}
= \frac{m(p)}{t'},
\]
and for every constant $c\geq1$ there is a sufficiently large $c'$ such that
\[
|C|=s\geq \frac{4ct'}{\eps'^2}\left(\log t'+\log\left(\frac{|B|}{\delta}\right)\right)
\in \frac{O(|B|)}{\eps'^2}\left(\log |B|+\log\left(\frac{1}{\delta}\right)\right).
\]

Substituting the query space $(P,w,\br{b},\zz)$, $d=1$, and $\delta/|B|$ instead of $\delta$ in Corollary~\ref{insofcor}, yields that with probability at least $1-\delta/(2|B|)$, we have
\begin{equation}\label{lastlast33}
\left|\sum_{p\in P}w(p)\zz(p,b)-\sum_{q\in C}u(q)\zz(q,b)\right|\leq \eps'. 
\end{equation}
Assume the event that~\eqref{lastlast3} holds for every $b\in B$ indeed occurs, which happens with probability at least $\delta/2$. Substituting the value of $\zz(p,b)$ from~\eqref{lastlast3} and multiplying by $\sum_{q\in P_b}w(q)$ yields
\begin{equation}\label{abcold}
\forall b\in B: \left|\sum_{p\in P_b}w(p)-\sum_{q\in C\cap P_b}u(q)\right|\leq \eps'\sum_{q\in P_b}w(q). 
\end{equation}
Hence,
\begin{align}
\label{abc}\sum_{b\in B}\pd(b,X)\left|\sum_{p\in P_b}w(p)-\sum_{p\in C\cap P_b}u(p)\right|
&\leq \eps'\sum_{b\in B} \pd(b,X)\sum_{q\in P_b}w(q)\\
\label{abc11}&\leq \eps'\sum_{b\in B} \rho(\alpha+1)\sum_{q\in P_b} w(q)g(q,X)\\
\nonumber&=\eps'\rho(\alpha+1) \sum_{q\in P} w(q)\pd(q,X),
\end{align}
where~\eqref{abc} is by~\eqref{abcold}, and~\eqref{abc11} is by the property of $(\alpha,\beta)$-assignment in~\eqref{secin2}.

Combining the previous inequalities all together yields the desired result
\begin{align}
\left|\sum_{p\in P}w(p)\pd(p,X)-\sum_{q\in C}u(q)\pd(q,X)\right|&\leq
 \eps'\sum_{p\in P}w(p)\pd(p,X)+\left|\sum_{q\in C\cup B}u(q)\pd(q,X)-\sum_{q\in C}u(q)\pd(p,X) \right|\label{comba}\\
&\leq \eps'\sum_{p\in P}w(p)\pd(p,X)+\sum_{b\in B}\pd(b,X)|\sum_{p\in P_b}w(p)-\sum_{p\in C\cap P_b}u(p)|\nonumber\\
&\leq \eps'(1+\rho(\alpha+1)) \sum_{q\in P} w(q)\pd(q,X)\label{combc}\\
&\leq \eps\sum_{q\in P} w(q)\pd(q,X),\nonumber
\end{align}
where~\eqref{comba} is by~\eqref{tri2}, and~\eqref{combc} is by~\eqref{abc11}.

Using the union bound on previous assumptions, this holds with probability at least $1-\delta/2-\delta/2=1-\delta$.
\end{proof}

\subsection{Positively weighted coresets}
In this section, we give a construction for a coreset that is guaranteed to output positive weights associated with each sampled point.

\begin{algorithm}[!htb]
\caption{$\coresetcond(P,w,\mathcal{B},s,\eps')$; See Theorem~\ref{mainthm25}\label{algmed5}}
\textbf{Input:} \quad\quad\= A weighted set $(P,w)$ where $w:P\to[0,\infty)$, $(\alpha,\beta)$-assignment $\mathcal{B}:P\to B$ for $(P,w,\CC,\pd)$, and sample size (integer) $s\geq 1$.\\
\textbf{Output:} A set $C\subseteq P$ and $u:C\times \CC(C)\to [0,\infty)$.

$(C\cup B,u):= \coresetc(P,w,\mathcal{B},s)\quad$\tcp{see Algorithm~\ref{algmed}}
\For{every $p\in \co$ and $\xx\in \CC(\co)$}
{\[
u'(p,\xx):=
\begin{cases}
u(p,\xx) & \text{if }\pd(\mathcal{B}(p),\xx)\leq \frac{ \pd(p,\mathcal{B}(p))}{\eps'}\\
0 & \text{otherwise}
\end{cases}.
\]}
\Return $(C\cup B,u')\quad$\tcp{$C\cup B$ is a multi-set}
\end{algorithm}

\begin{theorem}\label{mainthm25}
Consider the variables in Theorem~\ref{mainthm2}. 
Let
\[
  s\geq\frac{c't}{\eps^2}\left(d\log t+\log\left(\frac{1}{\delta}\right)  \right)
\]
and let $(C\cup B,u')$ be the output of a call to algorithm $\coresetcond(P,w,\mathcal{B},s,\eps')$; see Algorithm~\ref{algmed5}.
Then, $C\subseteq P$, $u:C\times \CC(C)\to[0,\infty)$, and with probability at least $1-\delta$, we have that for all $X\in \CC(C)$, 
\begin{align*}
\Bigg|\sum_{p\in P}w(p)\pd(p,X)-&\sum_{q\in C\cup B}u'(q,X)\pd(q,X)\Bigg|\leq \eps\sum_{p\in P}w(p)\pd(p,X).
\end{align*}
\end{theorem}
\begin{proof}
The proof is the same as the proof of Theorem~\ref{mainthm2} except for replacing $u(p)$ with $u'(p,\CC)$ everywhere, and replacing the bound on~\eqref{align1} by
\begin{align*}
\Bigg|\sum_{p\in F}\frac{v(p)(\pd(p,\xx)-\pd(p',\xx))}{\sum_{q\in P}w(q)\pd(q,\xx)}\Bigg|&\leq \frac{\eps}{4}\sum_{p\in F}\frac{(w(p)+u'(p,\xx))\pd(p',\xx)}{\sum_{q\in P}w(q)\pd(q',\xx)}\\
&=\frac{\eps}{4}\sum_{p\in F}\frac{w(p)\pd(p',\xx)}{\sum_{q\in P}w(q)\pd(q',\xx)}\leq \frac{\eps}{4}.
\end{align*}
where the first inequality is similar to~\eqref{align2}, and the equality is since $u'(p,\xx)=0$ for every
\[
p\in F=\br{p\in P\mid \pd(\mathcal{B}(p),\xx)>\frac{\pd(p,\mathcal{B}(p))}{\eps'}}.
\]
\end{proof}

\section{Empirical Evaluations}\label{emp}
This concludes our discussion of the general sensitivity sampling framework. 
Although our contribution is primarily theoretical, we nevertheless performed empirical evaluations in Python 3.6 via the Numpy and Scipy.sparse libraries on a desktop machine with an Intel i7-6850K CPU @ 3.60GHZ, 64GB RAM. 
We consider coreset constructions based on sensitivity sampling for bicriteria algorithms (Algorithm 1), general loss functions that satisfy the weak triangle inequality (Algorithm 2), and the conditional normalized distance (Algorithm 3). 
We compared Algorithms 1-3 to uniform sampling on $k$-means clustering on both relatively small offline data and large-scale streaming data that cannot fit into memory.
Algorithms 1-3 each require a bicriteria algorithm to approximate the importance of each point; we use \texttt{kmeans++} with $\alpha=O(\log k)$ and $\beta=1$ to approximate the importances, so that the runtime is linear.

\subsection{Evaluations on Offline Data}\label{smalldat}
 	\begin{figure*}[!htb]
		\begin{subfigure}{\sca\textwidth}
		\centering
		\includegraphics[scale=\sca]{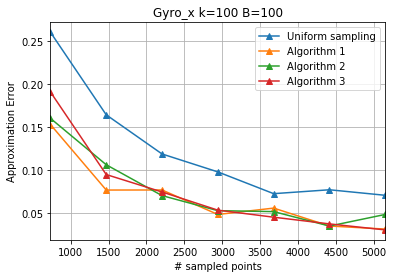}
		\caption{\label{klinemeansa} $x$ measure, $k=100$}
	\end{subfigure}
		\begin{subfigure}{\sca\textwidth}
		\centering
		\includegraphics[scale=\sca]{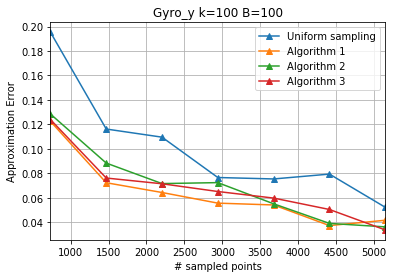}
		\caption{\label{klinemeansb} $y$ measure, $k=100$}
	\end{subfigure}
	\begin{subfigure}{\sca\textwidth}
		\centering
		\includegraphics[scale=\sca]{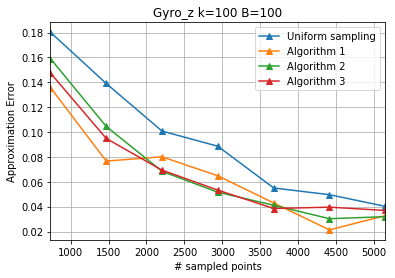}
		\caption{\label{klinemeansc} $z$ measure, $k=100$}
	\end{subfigure}

	\begin{subfigure}{\sca\textwidth}
		\centering
		\includegraphics[scale=\sca]{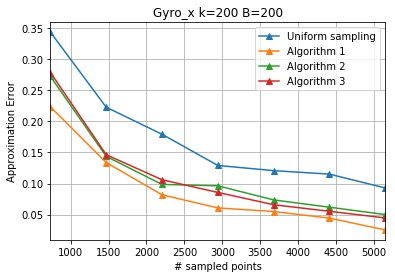}
		\caption{\label{klinemeansd} $x$ measure, $k=200$}
	\end{subfigure}
		\begin{subfigure}{\sca\textwidth}
		\centering
		\includegraphics[scale=\sca]{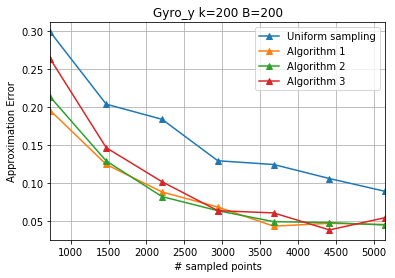}
		\caption{\label{klinemeanse} $y$ measure, $k=200$}
	\end{subfigure}
	\begin{subfigure}{\sca\textwidth}
		\centering
		\includegraphics[scale=\sca]{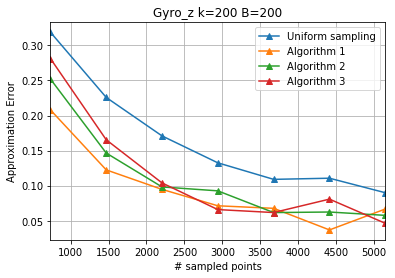}
		\caption{\label{klinemeansf} $z$ measure, $k=200$}
	\end{subfigure}

\caption{\small\label{F1} Experimental results on gyroscope data for uniform sample compared to our 3 algorithms.}
\end{figure*}

	\begin{figure*}[!htb]
		\begin{subfigure}[h]{\sca\textwidth}
		\centering
		\includegraphics[scale=\sca]{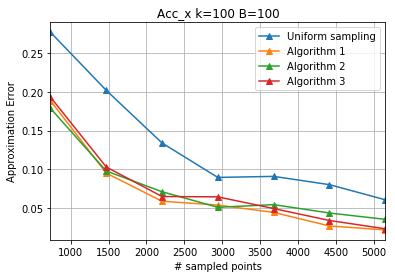}
		\caption{\label{klinemeansaa} $x$ measure, $k=100$}
	\end{subfigure}
		\begin{subfigure}[h]{\sca\textwidth}
		\centering
		\includegraphics[scale=\sca]{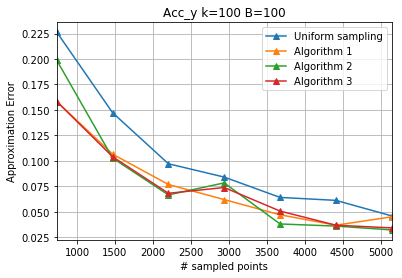}
		\caption{\label{klinemeansbb} $y$ measure, $k=100$}
	\end{subfigure}
	\begin{subfigure}[h]{\sca\textwidth}
		\centering
		\includegraphics[scale=\sca]{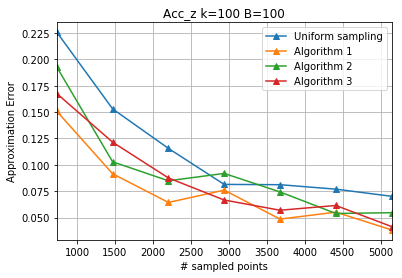}
		\caption{\label{klinemeanscc} $z$ measure, $k=100$}
	\end{subfigure}

	\begin{subfigure}[h]{\sca\textwidth}
		\centering
		\includegraphics[scale=\sca]{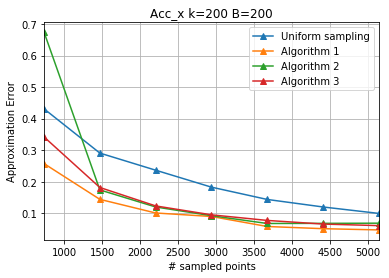}
		\caption{\label{klinemeansdd} $x$ measure, $k=200$}
	\end{subfigure}
		\begin{subfigure}[h]{\sca\textwidth}
		\centering
		\includegraphics[scale=\sca]{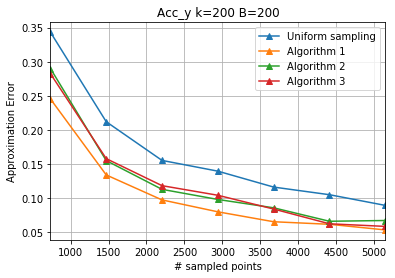}
		\caption{\label{klinemeansee} $y$ measure, $k=200$}
	\end{subfigure}
	\begin{subfigure}[h]{\sca\textwidth}
		\centering
		\includegraphics[scale=\sca]{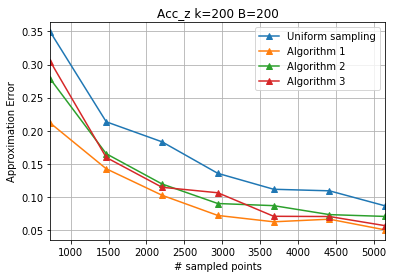}
		\caption{\label{klinemeansff} $z$ measure, $k=200$}
	\end{subfigure}

\caption{\small\label{F2} Experimental results on accelerometer data for uniform sample compared to our 3 algorithms. 
}
\end{figure*}
For experiments on small offline datasets, we compared our coreset constructions for $k$-means clustering in Algorithms 1-3 vs. uniform sampling on the datasets: (i) Gyroscope data and (ii) Accelerometer data. 
Collected by~\cite{anguita2013public}, and can be found on \cite{smart}, the experiments have been carried out with a group of 30 volunteers within an age bracket of 19-48 years. 
Each person performed six activities (walking, walking upstairs, walking downstairs, sitting, standing, laying) while wearing a Samsung Galaxy S II smartphone on the waist. 
Using its embedded gyroscope (resp. accelerometer), 3-axial angular velocity (resp. linear acceleration) were captured at a constant rate of 50Hz. 
The experiments have been video-recorded to label the data manually. 
Data was collected from $n=7352$ measurements; each instance consists of measurements from $d=3$ dimensions: $x$, $y$, $z$, each in a size of 128. 

We ran Algorithms 1-3 and uniform sampling on the above six datasets with different sample/coreset size, between 1000 to 7000, with $k=100$ and $k=200$.  
The multiplicative approximation error (empirical $\eps$) was calculated by $\eps:=\frac{\cost(A,Q_C)-\cost(A,Q_A)}{\cost(A,Q_A)}$, where $A$ is the matrix whose rows correspond to the $n$ input points, $Q_A$ corresponds to the $k$ centers of the whole data (two Lloyd's iterations after \texttt{kmeans++} initialization) and $Q_C$ is the clustering of the coreset. 
Our results show a significant improvement of our algorithms over uniform sampling; we present the gyroscope data evaluations in Figure~\ref{F1} and the accelerometer data evaluations in Figure~\ref{F2}.


\subsection{Evaluations on Streaming Data}
To handle large-scale streaming data that cannot fit into memory, our system separates the $n$ points of the data into chunks of a desired size of coreset, called $m$. 
We use a merge-and-reduce framework on a binary tree, e.g.~\cite{feldman2010coresets}, where each node is a coreset of the union of the data represented by its children nodes and the bottom layer of the tree consists of consecutive chunks of the data of size $4516$. 
Thus the root of the tree is a coreset of the whole data. 
We build a tree of height $10$ for our data, dividing the $n=4624611$ input points across $1024$ chunks of size $4516$. 

\begin{figure*}[!htb]
	\begin{subfigure}{0.5\textwidth}
		\centering
		\includegraphics[scale=0.5]{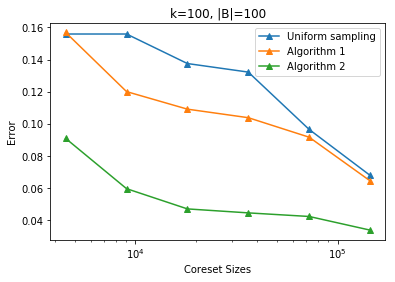}
		\caption{\label{bd1} $k=100$}
	\end{subfigure}
	\begin{subfigure}{0.5\textwidth}
		\centering
		\includegraphics[scale=0.5]{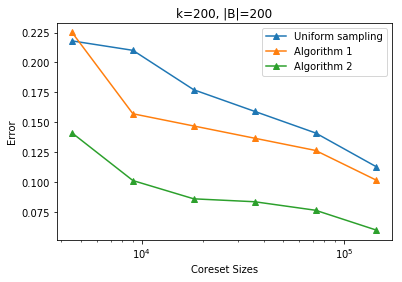}
		\caption{\label{bd2} $k=200$}
	\end{subfigure}
\caption{Experimental results for $k$-means clustering on a data stream of a created document-term matrix of Wikipedia, with $k=100$ in \ref{bd1} and $k=200$ in \ref{bd2}. 
Algorithms 1 and 2 in figures based on general sensitivity sampling for bicriteria algorithms and conditional normalized distances, respectively.}
\label{F3}
\end{figure*}

\paragraph{Wikipedia Dataset.}
We compared uniform sampling to Algorithms 1 and 3 for $k$-means clustering on a created document-term matrix of Wikipedia (parsed enwiki-latest-pages-articles.xml.bz2-rss.xml from \cite{wic2019}), i.e. sparse matrix with 4624611 rows and 100k columns where each cell $(i,j)$ equals the value of how many appearances the word number $j$ has in article number $i$. 
We use a standard dictionary of the 100k most common words in Wikipedia \cite{dic2012}.
We concatenated the coreset received in each floor and compared the received error in each floor. 
The error we determined was calculated by the formula $\frac{\cost(A,Q_C)-\cost(A,Q_A)}{\cost(A,Q_A)}$, where $A$ is the original data matrix, $Q_A$ is the clustering of the whole data (Lloyd's iterations until 1\% convergence, after ++ initialization) and $Q_C$ is the clustering of the coreset. 
We used two values of $k$, 100 and 200. 
We present our results in Figure \ref{F3}.
We concatenated the coreset received in each floor and compared the received error in each floor. 
The error we determined was calculated by the formula $\frac{\cost(A,Q_C)-\cost(A,Q_A)}{\cost(A,Q_A)}$, where $A$ is the original data matrix, $Q_A$ is the clustering of the whole data (Lloyd's iterations until 1\% convergence, after ++ initialization) and $Q_C$ is the clustering of the coreset. 
We present our results in Figure \ref{F3} for $k=100$ and $k=200$. 
Similar to the offline evaluations, we obtain better results for our algorithms than uniform sampling. 
However, unlike than the offline data, here the conditional normalized algorithm gets much better results than the general sensitivity sampling algorithm.

\paragraph{Algorithms.}
The algorithms we compared are uniform sampling, and our Algorithm \ref{one} and \ref{algmed5}. 

\paragraph{Results. }
We concatenated the coreset received in each floor and compared the received error in each floor. 
The error we determined was calculated by the formula $\frac{\cost(A,Q_A)-\cost(A,Q_C)}{\cost(A,Q_A)}$, where $A$ is the original data matrix, $Q_A$ is the clustering of the whole data (Lloyd's iterations until 1\% convergence, after ++ initialization) and $Q_C$ is the clustering of the coreset. 
We used two values of $k$, 100 and 200. 
We present our results in Figure \ref{F3}.

\paragraph{Discussion. }
Indeed also for this dataset we got better results for our algorithm than uniform sampling. However, unlike than in Section~\ref{emp}, here Algorithm 2 gets much better results than Algorithm 1.
\bibliographystyle{alpha}
\bibliography{references.bib}
\appendix
\end{document}